\documentclass[11pt]{article}
\usepackage{graphicx,amsmath,amssymb}
\usepackage{xspace,hyperref,cite}
\newcommand{\cinst}[2]{$^{\mathrm{#1)}}$~#2\par}
\newcommand{\crefi}[1]{$^{\mathrm{#1)}}$}
\newcommand{\HRule}{\rule{0.4\linewidth}{0.3mm}}

\newcommand{\pt}{\ensuremath{p_\mathrm{T}}\xspace}
\newcommand{\jpsi}{\ensuremath{\mathrm{J}\hspace{-.08em}/\hspace{-.14em}\psi}\xspace}
\newcommand{\QQbar}{\ensuremath{Q \overline{Q}}\xspace}
\newcommand{\ccbar}{\ensuremath{c \overline{c}}\xspace}
\newcommand{\ppbar}{\ensuremath{\mathrm{p} \overline{\mathrm{p}}}\xspace}
\newcommand{\lth}{\ensuremath{\lambda_\vartheta}\xspace}
\newcommand{\lph}{\ensuremath{\lambda_\varphi}\xspace}
\newcommand{\ltp}{\ensuremath{\lambda_{\vartheta\varphi}}\xspace}
\newcommand{\ltilde}{\ensuremath{\,\tilde{\!\lambda}}\xspace}

\makeatletter
\newcommand{\lrarrow}{\mathrel{\mathpalette\lrarrow@\relax}}
\newcommand{\lrarrow@}[2]{%
  \vcenter{\hbox{\ooalign{%
    $\m@th#1\mkern6mu\rightarrow$\cr
    \noalign{\vskip2.5pt}
    $\m@th#1\leftarrow\mkern6mu$\cr
  }}}%
}
\makeatother

\begin{document}

\begingroup
\thispagestyle{empty} \baselineskip=14pt
\parskip 0pt plus 5pt

\begin{center}
{\Large\bf \boldmath 
On the polarization of the non-prompt contribution\\[3mm]
to inclusive \jpsi production in pp collisions}

\bigskip\bigskip

Pietro Faccioli\crefi{1} and Carlos Louren\c{c}o\crefi{2}

\bigskip
\textbf{Abstract}

\end{center}

\begingroup
\leftskip=0.4cm \rightskip=0.4cm
\parindent=0.pt

Of the J/$\psi$ mesons (inclusively) produced in pp collisions, a big fraction results from B decays, increasing with transverse momentum and exceeding 50\% for $p_{\rm T} > 20$~GeV. These events must be subtracted in measurements of the polarization of prompt J/$\psi$ mesons. While several studies have addressed the $\psi$(2S) and $\chi_c$ impact on the determination of the polarization of the directly-produced J/$\psi$ mesons, the theoretical and experimental knowledge of the non-prompt polarization is very poor. Furthermore, non-prompt J/$\psi$ polarization measurements can provide interesting information on quarkonium hadroproduction, complementing the studies of prompt production. We review the method of measuring the polarization of non-prompt J/$\psi$ mesons (produced in decays of unreconstructed B mesons and detected in the dilepton channel), in conditions typical of LHC experiments studying J/$\psi$ production. Realistic model-independent scenarios are validated with data from experiments studying $e^+e^- \to \Upsilon$(4S) interactions, converted to the high-momentum regime using B differential cross sections measured at the LHC. The non-prompt J/$\psi$ polarization measurements are seen to remain dependent on the event selection criteria, even after correcting for the dilepton acceptance and efficiencies. This implies that reproducible definitions of all relevant analysis choices must be reported with the polarization result, for rigorous comparisons with other measurements and/or theoretical calculations. We also discuss how the non-prompt J/$\psi$ polarization significantly depends on the relative importance of two complementary $\mathrm{B}\to {\rm J}/\psi$ decay topologies, two-body (reasonably dominated by singlet production) and multi-body (including octet contributions), providing, hence, valuable information for studies of the charmonium formation mechanisms.

\endgroup

\vfill
\begin{flushleft}
\HRule\\

\cinst{1} {LIP, Lisbon, Portugal, Pietro.Faccioli@cern.ch} 
\cinst{2} {CERN, Geneva, Switzerland, Carlos.Lourenco@cern.ch}

\end{flushleft}
\endgroup

\newpage

\section{Introduction}
\label{sec:intro}

Quarkonium production studies provide crucial information 
on the mechanisms underlying hadron formation~\cite{Brambilla:2011}. 
The colour singlet model~\cite{Baier:1983}, 
where the quarkonium can only be formed as an already colour neutral ${}^3S_{\,1}^{[1]}$ state, 
has been seen to not reproduce the cross sections and polarizations 
of quarkonia produced at midrapidity in high-energy hadron collisions, 
both at the Tevatron~\cite{CDF:psi,CDF:psi-pol,CDF:Upsilon-pol,D0:Upsilon-pol} 
and at the LHC~\cite{CMS:psi,CMS:YnS,ATLAS:psi2S,ATLAS:YnS,CMS:psi-pol,CMS:Upsilon-pol}.
After roughly two decades of ambiguous and inconsistent theory-data comparisons,
leading to puzzling interpretations of quarkonium polarization data~\cite{Faccioli:EPJC69}, 
it has recently been shown that the NRQCD framework~\cite{NRQCD}, 
which includes quarkonium production through intermediate colour octet \QQbar states,
is able to describe, consistently and simultaneously,
the \emph{prompt} quarkonium cross sections and polarizations measured by ATLAS and 
CMS~\cite{CMS:psi,CMS:YnS,ATLAS:psi2S,ATLAS:YnS,CMS:psi-pol,CMS:Upsilon-pol},
while suggesting that production via one specific octet state, 
the unpolarized ${}^1S_{\,0}^{[8]}$, 
dominates over all other processes, 
at least in the midrapidity and high transverse momentum (\pt) domain covered by these 
experiments~\cite{Faccioli:PLB736,Faccioli:PLB773,Bodwin:2015,Faccioli:EPJC78p268}.
A crucial ingredient of this conclusion is the thought-provoking~\cite{Faccioli:2019} 
unpolarized prompt \jpsi and $\psi$(2S) production observed 
by CMS~\cite{CMS:psi-pol} in the midrapidity region and also 
by LHCb~\cite{LHCb:psi-pol, LHCb:psip-pol} at forward angles.

Besides the promptly-produced \jpsi mesons 
(which include both the mesons directly produced from the partonic interaction 
and those resulting from ``feed-down decays'' of heavier charmonia),
a significant fraction of the \jpsi mesons detected in high energy experiments 
comes from decays of B hadrons (mostly from ${\rm B}^{\pm}$ and ${\rm B}^0$ decays).
They are commonly known as \emph{non-prompt} mesons 
and are characterized by an exponential ``lifetime distribution'',
measured from the distance between the production vertex (the pp collision point)
and the decay vertex (where the \jpsi is produced and immediately decays to a pair of muons or electrons).
At the LHC, the fraction of non-prompt \jpsi mesons increases 
from about 10\% at very low \pt to around 70\% 
for $\pt > 50$\,GeV~\cite{ALICE-Bfraction-vs-pT,CMS-Bfraction-vs-pT,ATLAS-Bfraction-vs-pT}.
The polarization of these mesons is conceptually and effectively different from that of the prompt ones.

Some publications report measurements of the polarization of an inclusive sample of \jpsi mesons, 
without subtracting the non-prompt ``background''~\cite{ALICE:Jpsi-pol-7TeV,ALICE:Jpsi-pol-8TeV}.
The non-negligible impact of that component, which, furthermore, significantly depends on \pt,
limits the accuracy that can be achieved in comparisons of such measurements 
with theory calculations (or with other measurements).
So far, the polarization of non-prompt \jpsi mesons produced in hadron collisions
has only been reported by one experiment, CDF, 
which used a (small) sample of \ppbar collisions at $\sqrt{s} = 1.8$\,TeV~\cite{CDF:nonprompt}.
Furthermore, this measurement suffers from rather large uncertainties. 
The CMS and LHCb experiments have shown that they can provide 
high-precision quarkonium polarization 
measurements~\cite{CMS:psi-pol,CMS:Upsilon-pol,LHCb:psi-pol,LHCb:psip-pol},
benefiting from very good measurement resolutions, signal-to-background ratios, 
and large event samples.
They could certainly obtain high-quality results for the polarizations of the non-prompt 
\jpsi and $\psi$(2S) states, 
which would provide very relevant information to understand quarkonium production
(how the heavy quark-antiquark pair binds into the final-state hadron),
given that non-prompt production reflects a complementary and independent interplay 
between the singlet and octet channels, with respect to prompt production. 
In this context, it is worth emphasising that polarization is a particularly discerning observable. 
In fact, while the differential non-prompt \jpsi cross section 
only reflects the production mechanism of the B mesons,
the corresponding \jpsi polarization probes the underlying quarkonium formation mechanism.

In this article we review the analysis methodology of a non-prompt \jpsi polarization measurement.
Samples of non-prompt \jpsi events are selected by exploiting the fact that 
the distance between the point where the parent B is produced 
(the proton-proton or proton-antiproton interaction point, say, usually called primary vertex) 
and the point where the \jpsi is produced and immediately decays (the dilepton vertex) 
is significantly larger than the uncertainty in the measurement of that distance.
This method is justified by the relatively large average decay length of the B mesons 
(of order 500~$\mu$m) 
with respect to the measurement resolution of most modern experiments, of around 10~$\mu$m. 
The \jpsi polarization can then be measured using the selected events, 
by fitting the dilepton angular distribution in the \jpsi rest frame, 
considered with respect to the same laboratory-referred directions 
as used in prompt polarization measurements, 
given that no information on the parent B mesons has been recorded.
We will focus on the high-momentum conditions typical of the LHC experiments.
The only physical inputs used in our analysis are the momentum spectrum and the polarization
of \jpsi samples produced in decays of B mesons, 
themselves produced in $e^+e^-$ collisions at the $\Upsilon$(4S) resonance,
as reported by CLEO~\cite{CLEO} and BaBar~\cite{BaBar},
as well as the \pt distribution of B mesons measured by 
CDF~\cite{CDF:Bplus-xsection}, 
ATLAS~\cite{ATLAS:Bplus-xsection-7TeV} and 
CMS~\cite{CMS:Bplus-xsection-7TeV}.

To understand how the analysis of the \jpsi dilepton decay distribution
is affected by the integration of the angular degrees of freedom of the (unobserved) B decay, 
we start by describing in detail the angular distribution of the full two-step decay, 
$\mathrm{B} \to \jpsi \, X$ followed by $\jpsi \to \ell^+ \ell^-$ (Section~\ref{sec:angulardistr}),
and also the momentum relations between mother and daughter particles (Section~\ref{sec:kinematics}).
Using the specific two-body decay $\mathrm{B} \to \jpsi \, \mathrm{K}$ as template, 
we then discuss how the very act of performing the analysis as a function of the \jpsi momentum 
(instead of using the B momentum) 
distorts the natural decay distribution in an unrecoverable way (Section~\ref{sec:observation}).
Another shaping effect is caused by the event selection requirements on the momenta of the leptons.
We will see in Section~\ref{sec:selections} that this effect does not disappear after 
the application of the acceptance and efficiency corrections evaluated in the usual angular analyses,
which only consider the dilepton degrees of freedom 
and necessarily ignore the unobserved B production and decay kinematics.
We will then (Section~\ref{sec:nonprompt}) translate the indications obtained in $e^+e^-$ collisions 
about topologies and resulting \jpsi polarizations of the contributing B decays 
into corresponding predictions for what an LHC experiment should observe 
in the presence of a similar mixture of B decays, 
or in extreme cases where individual categories of topologies would prevail.
Section~\ref{sec:summary} summarizes the article,
emphasizing how, for this kind of measurements,
comparisons between several experiments and/or with theory predictions require particular care.

\section{The $\mathrm{B} \to \jpsi \to \ell^+ \ell^-$ angular distribution}
\label{sec:angulardistr}

In this section we study the two-step decay $\mathrm{B} \to \jpsi \, X$, $\jpsi \to \ell^+ \ell^-$,
where $X$ is an accompanying object, such as a kaon.
The process has four degrees of freedom, 
represented by the angles $\Theta$ and $\Phi$,
describing the direction of the \jpsi in the B rest frame, 
and $\vartheta$ and $\varphi$, 
the (positive) lepton emission angles in the \jpsi rest frame.

\begin{figure}[t]
\centering
\resizebox{0.9\linewidth}{!}{\includegraphics{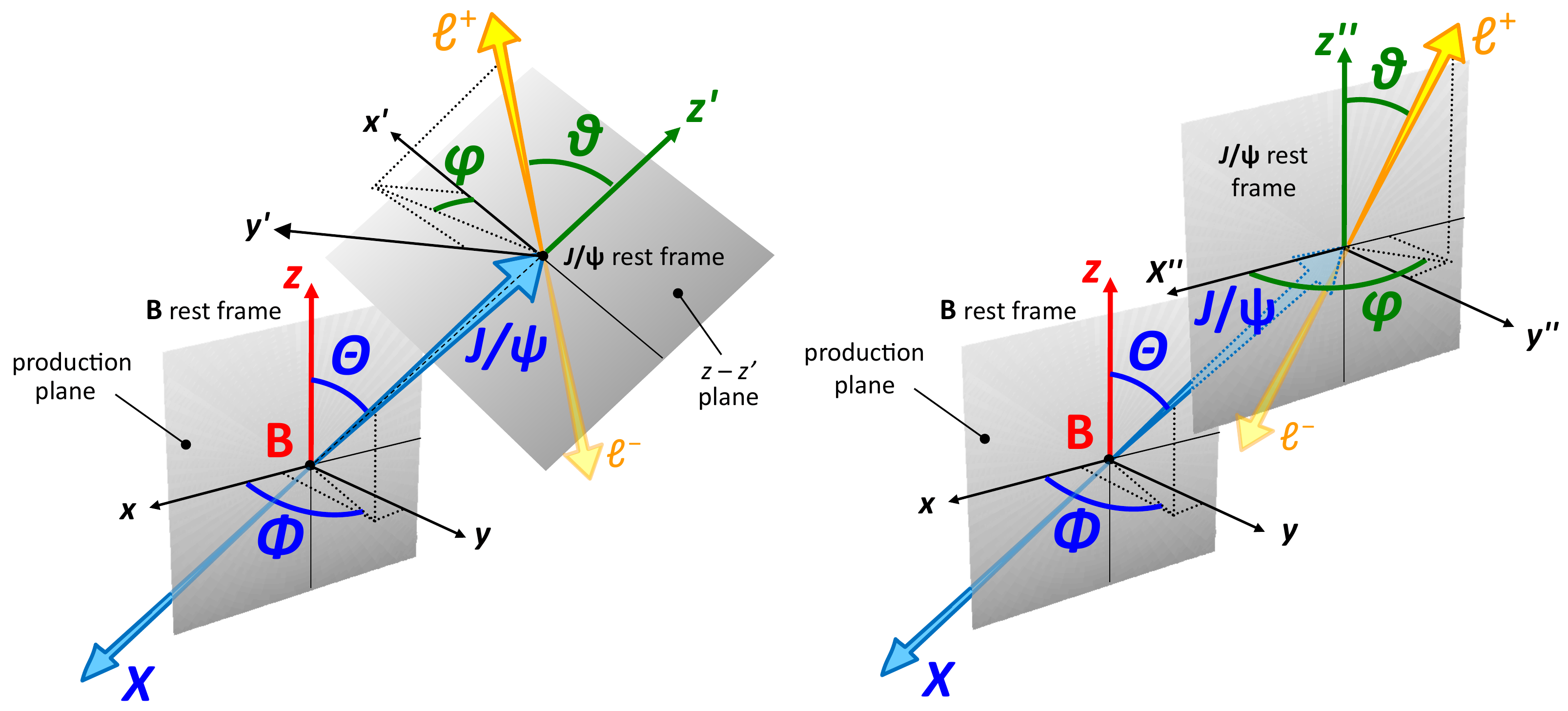}}
\caption[]{Two alternative definitions of the angles used in the description of the cascade decay
$\mathrm{B} \to \jpsi \, X$, $\jpsi \to \ell^+ \ell^-$.
The $\Theta$ and $\Phi$ angles are always measured in the $\mathrm{B}$ polarization frame ($x,y,z$), 
for example the HX frame,
while the (positive) lepton emission angles in the \jpsi rest frame, $\vartheta$ and $\varphi$,
are either defined with respect to the ($x',y',z'$) \emph{cascade helicity} (cHX) system of axes,
as shown in the left diagram, 
or with respect to a system of axes geometrically identical to $x,y,z$,
the \emph{cloned cascade frame} (CC),
as shown in the right diagram.}
\label{fig:J0_decay_frames}
\end{figure}

Figure~\ref{fig:J0_decay_frames} shows two alternative definitions of these variables. 
The angles $\Theta$ and $\Phi$ are defined with respect to the polarization frame chosen for the B mesons, 
such as the centre-of-mass helicity frame, HX, 
which has as $z$ axis the B direction in the laboratory, 
or the Collins--Soper frame, CS, 
where $z$ is the average of the momentum directions of the two colliding hadrons in the B rest frame; 
in both cases, the $zx$ plane coincides with the B production plane. 

For the dilepton decay, the left diagram of Fig.~\ref{fig:J0_decay_frames} 
shows the \emph{cascade helicity frame} (cHX),
where the polarization axis is the \jpsi direction in the B rest frame and
the azimuthal angle is measured in the plane containing 
the polarization axes of the two particles ($z$ and $z'$).
The B meson has zero angular momentum ($J=0$) and emits its products isotropically,
so that the angular variables $\cos\Theta$ and $\Phi$ are uniformly distributed.
The system $\jpsi + X$ has angular momentum $J = 0$ and, hence, 
projection $J_z = 0$ on any $z$ axis, so that, in general,
\begin{equation}
J_z^{\jpsi} + J_z^{X} + I_z^{\jpsi\text{-}X} = 0 \, .
\label{eq:J0decay_Jz}
\end{equation}
This relation includes a possible orbital angular momentum component $\vec{I}^{\jpsi\text{-}X}$
between the two final particles.
In the two-body decay $\mathrm{B}\, (J=0) \to \jpsi\, (J=1) \; \mathrm{K}\, (J=0)$, for example, 
we know that $I_z^{\jpsi\text{-}\mathrm{K}} = 1$, to ensure angular momentum conservation.
Along the cHX axis $z'$,
defined by the common direction of the back-to-back \jpsi and $X$ momenta, 
$I_{z'}^{\jpsi\text{-}X}$ vanishes because $\vec{I}^{\jpsi\text{-}X}$ is perpendicular to the linear momenta: 
only the individual spins of the \jpsi and $X$ have to be considered in the projected sum. 
The component $J_{z'}^{X}$ is well defined, and equal to 0, 
when $X$ is a kaon or another $J=0$ particle, in which case $J_{z'}^{\jpsi} = 0$: 
the \jpsi has an intrinsic longitudinal polarization.
The four-dimensional angular distribution becomes
\begin{equation}
W_{\mathrm{cHX}}(\cos\Theta, \Phi, \cos\vartheta, \varphi) \, \propto \, 1 + \lambda_0 \cos^2 \vartheta \, ,
\label{eq:J0decay_cHX_distr}
\end{equation}
where the \emph{natural} polarization, $\lambda_0$, is $-1$ if $X$ is a \mbox{$J=0$} particle, 
and there is no dependence on $\Theta$ and $\Phi$.
To measure this distribution, an experiment must reconstruct not only the \jpsi but also the B meson, 
to determine the momentum and rest frame of B, 
which are needed for the definition of the cHX polarization axis.

We will now consider the common case (relevant for the present study)
where the \jpsi polarization measurement ignores (i.e., implicitly integrates out) 
the degrees of freedom of $X$ and observes only the two lepton tracks, 
in a frame defined only using the momentum directions of the colliding hadrons, such as the \jpsi HX frame. 
The measured distribution is, in this case, very different from Eq.~\ref{eq:J0decay_cHX_distr}, 
because the dilepton variables $\cos\vartheta$ and $\varphi$, 
as measured in the HX frame, 
have no definite correlation to those defining the natural decay distribution as observable in the cHX frame. 
Actually, the two sets of angular variables tend to be \emph{fully uncorrelated}.
In fact, the HX frame is defined with respect to directions that are fixed in the laboratory, 
while the cHX frame uses the direction of the \jpsi in the B rest frame, 
which is generated following a spherical distribution: 
the $\cos\Theta$ and $\Phi$ angles are uniformly distributed. 
According to this reasoning, a \emph{fully} unpolarized \jpsi should be observed 
when the angular measurement treats the \jpsi as if it were directly produced.

We will see now that, however, 
the measurement itself perturbs the spherical distribution naturally produced by the decay of a $J = 0$ particle,
leading to the observation of a more or less \emph{anisotropic} dilepton distribution.
In fact, unavoidable experimental selections sculpt the 
$(\cos\Theta, \Phi)$ two-dimensional distribution, which loses its uniformity.
To evaluate these experimental effects, it is important to analyze 
the correlation between the B decay angles and those of the dilepton decay in the HX frame,
which is not feasible using Eq.~\ref{eq:J0decay_cHX_distr}, 
where the four-dimensional distribution $W$ is independent of $\cos\Theta$ and $\Phi$.
Therefore, it is convenient to use another definition of the \jpsi polarization frame,
represented by the right diagram of Fig.~\ref{fig:J0_decay_frames}.
While the $x, y, z$ axes, 
used for the measurement of the $\Theta$ and $\Phi$ emission angles  of the \jpsi,
remain the same as in the previous definition
(e.g., $z$ is the polarization axis in the B HX frame),
the axes used for the dilepton decay in the $\jpsi$ rest frame, $x'', y'', z''$, 
are now exact geometrical clones of the $x, y, z$ axes,
obtained by a simple translation,
not involving any Lorentz boosts of the physical references.
In practice, the dimensionless unit vectors of the $x, y, z$ axes,
defined in the B rest frame (the B kinematics must be reconstructed),
are used, identically, as unit vectors of the $x'', y'', z''$ axes in the $\jpsi$ rest frame.
This choice, here referred to as the \emph{cloned cascade frame} (CC),
might seem physically abstract and perhaps counter-intuitive.
There is a limit, however, 
where the $x'', y'', z''$ axes simply reduce to those of the ``ordinary'' HX frame
(or any other $x, y, z$ frame) of the \jpsi, 
i.e., the one defined in terms of beam directions Lorentz-boosted to the \jpsi rest frame:
when the B and \jpsi laboratory momenta are much larger than their mass difference.
In that limit, the directions of, say, 
the HX axis in the B rest frame and the HX axis in the \jpsi rest frame tend to coincide
(as will be quantitatively described in Section~\ref{sec:kinematics}).

We note that the definition of the CC frame requires the specification of the frame used
for the B decay, i.e., which exact frame is being ``cloned''; 
we can refer to, for example, the HX CC or CS CC frames.
In the remainder of this article, if this specification is absent it is implicit
that we are using the HX frame as ``master'' frame.
For a decay of the kind \mbox{$\mathrm{B} \to \jpsi \, X$},
given the relatively small difference between the B and \jpsi masses, 
in comparison to their typical laboratory momenta,
this condition is satisfied in most of the kinematic domains of the LHC experiments.
In this limit, the determination of the $x'', y'', z''$ axes decouples from the knowledge of the B momentum,
so that the polarization measurement in the CC frame can effectively be performed
without observing the accompanying particle $X$ 
(and without reconstructing the B rest frame).

\begin{figure}[t]
\centering
\resizebox{0.75\linewidth}{!}{\includegraphics{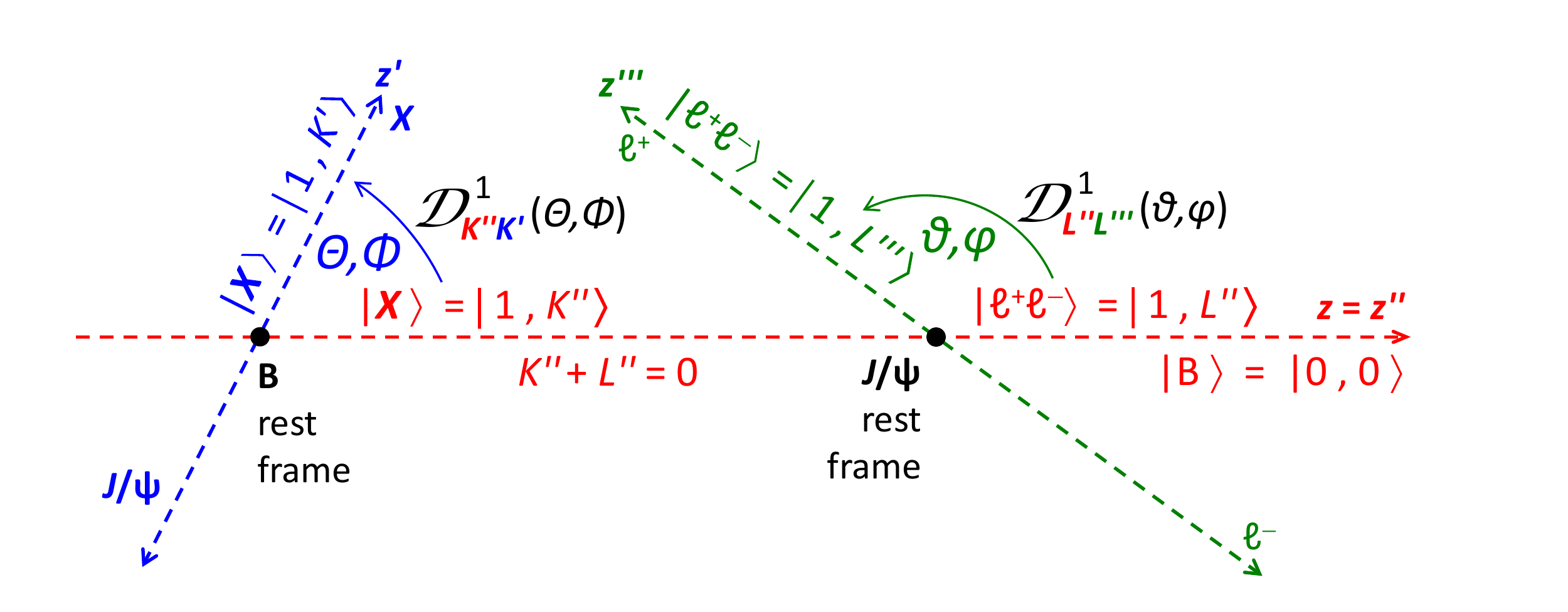}}
\caption[]{Sketch of the cascade decay 
$\mathrm{B} \to \jpsi \, X$, $\jpsi \to \ell^+ \ell^-$,
indicating the axes (CC frame), the decay angles,
the angular momentum states of the involved particles,
and the Wigner matrix elements used for their rotation.
The red line represents the common direction of the B and \jpsi polarization axes
($z$ and $z''$, respectively) in the CC frame.}
\label{fig:cascade_notations}
\end{figure}

To determine the new expression of the four-dimensional angular distribution,
we start by writing the amplitude of the two-body decay process $\mathrm{B} \to \jpsi \, X$.
Figure~\ref{fig:cascade_notations} summarizes the notations used 
for the angular momentum states of the involved particles, the axes, and their rotations.
As mentioned above, in the considered case, 
$X$ has a definite angular momentum projection, $K'=0$, 
along the $z'$ (cHX) axis.
There is also an orbital momentum component that now, 
with respect to the CC polarization axis $z''$, 
we cannot ignore.
In order to use simple two-body angular momentum sum rules,
we attribute the orbital angular momentum to $X$.
In practice, we consider $X$ as a state that has, whatever its identity, 
total angular momentum $J=1$, including the orbital part, so that, 
when we add it to the (also $J=1$) \jpsi,
we can recover the zero angular momentum of the B mother particle.

The Wigner matrix needed to rotate the angular momentum of $X$
from the $x',y',z'$ axes to the $x'', y'', z''$ axes is, therefore,
$\mathcal{D}_{K'' \, K'}^{1}(\Theta, \Phi)$,
where $K''$ is the $J_{z''}$ projection of $X$ on the $z''$ axis,
\begin{equation}
|X; 1, K' \rangle_{z'} =
\sum_{K'' = 0, \pm 1} \mathcal{D}_{K'' \, K'}^{1}(\Theta, \Phi) \;\; |X; 1, K'' \rangle_{z''} \, .
\label{eq:rotate_X_cHX_to_CC}
\end{equation}
Since we only consider parity-conserving terms of the decay distribution,
$\Theta$ and $\Phi$ indifferently denote the \jpsi or $X$ directions,
while if we wanted to obtain correct signs for the parity-violating terms,
the Wigner matrix for the rotation of $X$ would be
$\mathcal{D}_{K'' \, K'}^{1}(\pi - \Theta, \pi + \Phi)$
(using the \jpsi direction to define $\Theta$ and $\Phi$,
as represented in Fig.~\ref{fig:J0_decay_frames}).
Indicating with $L''$ a generic $J_{z''}$ projection of the \jpsi on $z''$,
the decay amplitude is given by
\begin{equation}
\begin{split}
\mathcal{A}(\mathrm{B} \to \jpsi_{L''} + X_{K'} ) 
& = \sum_{K'' = 0, \pm 1} \!_{z''}\langle \jpsi \, X ; 1, L'', 1, K'' 
\, | \, \mathcal{B} \, | \, \mathrm{B}; 0,0 \rangle_{z''}
\, \mathcal{D}_{K'' \, K'}^{1*}(\Theta, \Phi) \\
& = \langle  1, L'', \, 1, -L'' \, | \,  0,0 \rangle \,\, \mathcal{D}_{-L'' \, K'}^{1*}(\Theta, \Phi) \; ,
\end{split}
\label{eq:S_to_VX_amplitude}
\end{equation}
where the operator $\mathcal{B}$, containing the dynamics of the decay,
can, in general, impose relations between the angular momentum states
of the B, \jpsi, and $X$ particles.

In the cases here considered, the relevant physical constraints derive from angular momentum conservation.
First, along $z''$, the \jpsi and $X$ particles must have opposite angular momentum projections,
being the daughters of a $J = 0$ state,
as expressed in the relation used in the second equality above:
\begin{equation}
{}_{z''}\langle \jpsi \, X ; \, 1, L'', \, 1,K'' \, | \, \mathcal{B} \, |\, \mathrm{B}; 0,0 \rangle_{z''}
\, \propto \, \delta_{K'',-L''} \, \langle 1, L'', \, 1, -L'' \, | \, 0,0 \rangle \, .
\label{eq:S_to_VX_J_eq_0_constraint}
\end{equation}
Second, in the specific case where $X$ is a $J=0$ particle (e.g., a kaon, a pion, or an $\eta$ meson)
angular momentum conservation effectively determines a definite natural longitudinal polarization for the \jpsi.
The Clebsch--Gordan coefficient $\langle  1, L'', \, 1, -L'' \, | \,  0,0 \rangle$
is $\sqrt{3}/3$ and $-\sqrt{3}/3$ for $L'' = 0$ and $\pm 1$, respectively.
The amplitude of the two-step decay process can then be written by including 
a factor expressing the rotation of the dilepton angular momentum state,
which has projection $L''' = \pm 1$ along its own flight direction in the \jpsi rest frame ($z'''$ axis),
onto the $z''$ axis (where it has projection identical to the \jpsi one, $L''$),
and summing over the possible $L''$ components of the \jpsi:
\begin{equation}
\begin{split}
& \mathcal{A}\,[\,\mathcal{B} \to \jpsi + X_{K'}, \, \jpsi \to (\ell^+ \ell^-)_{L'''} ] \propto \\
\sum_{L'' = 0, \pm 1} \langle & 1, L'', 1, -L'' \, | \, 0,0 \rangle \;
\mathcal{D}_{-L'' \, K'}^{1*}(\Theta, \Phi) \; 
\mathcal{D}_{L'' \, L'''}^{1*}(\vartheta, \varphi) \; .
\end{split}
\label{eq:S_to_VX_V_to_ll_amplitude}
\end{equation}

The final expression of the angular distribution is obtained by squaring
Eq.~\ref{eq:S_to_VX_V_to_ll_amplitude}, and summing over $L'''=\pm 1$
and over the relevant $K'$ values, which depend on the identity of $X$.
Among the cases that we will consider,
\mbox{$K'=0$} if $X$ is a kaon (or any other $J=0$ particle), 
implying that the \jpsi has a fully longitudinal polarization in the cHX frame,
\mbox{$\lambda_0 = -1$}.
The resulting distribution for a generic natural polarization $\lambda_0$ is:
\begin{equation}
\begin{split}
W_{\mathrm{CC}}&(\cos\Theta, \Phi, \cos\vartheta, \varphi) \propto \frac{1}{3+\lambda_0} \; \Big[ \, 2 \,
+ \lambda_0 \, (1 -  \cos^2\Theta - \cos^2\vartheta + 3 \cos^2\Theta \, \cos^2\vartheta \, ) \\
& + \lambda_0 \, \sin^2\Theta \, \sin^2\vartheta \, \cos 2 (\varphi-\Phi) \,
+ \lambda_0 \, \sin 2 \Theta \, \sin 2 \vartheta \, \cos \, (\varphi-\Phi) \, \Big] \, .
\end{split}
\label{eq:J0decay_CC_distr}
\end{equation}

This expression is fully symmetric with respect to an exchange between
the B and \jpsi decay angles, 
\mbox{$(\Theta, \Phi) \lrarrow (\vartheta, \varphi)$},
and, moreover, only depends on the two azimuthal angles through their difference, 
$\varphi - \Phi$.
We can, however, rewrite it so as to give emphasis to the dilepton part,
obtaining the same angular distribution as used in 
prompt \jpsi polarization measurements~\cite{Faccioli:EPJC69},
with anisotropy parameters that only depend on $\cos\Theta$ and $\Phi$:
\begin{equation}
\begin{split}
\lth & = \frac{ -\lambda_0 \, ( 1 - 3 \cos^2 \Theta ) }{ 2 + \lambda_0 \, ( 1 - \cos^2 \Theta ) } \, , \\
\lph & = \frac{ \lambda_0 \, \sin^2 \Theta \, \cos 2 \Phi }{ 2 + \lambda_0 \, ( 1 - \cos^2 \Theta ) } \, , \quad\quad
\lph^\bot = \frac{ \lambda_0 \, \sin^2 \Theta \, \sin 2 \Phi }{ 2 + \lambda_0 \, ( 1 - \cos^2 \Theta ) } \, , \\
\ltp & = \frac{ \lambda_0 \, \sin 2 \Theta \, \cos \Phi }{ 2 + \lambda_0 \, ( 1 - \cos^2 \Theta ) }  \, , \quad\quad
\ltp^\bot = \frac{ \lambda_0 \, \sin 2 \Theta \, \sin \Phi }{ 2 + \lambda_0 \, ( 1 - \cos^2 \Theta ) } \, .
\end{split}
\label{eq:J0decay_CC_distr_lambdas}
\end{equation}

Clearly, the distribution becomes isotropic if \mbox{$\lambda_0 = 0$}.
We can also recognize from Eq.~\ref{eq:J0decay_CC_distr} that 
the average over a uniform (or linear) $\cos\vartheta$ distribution 
(giving \mbox{$\langle \cos^2\vartheta\rangle = 1/3$}) 
and over the azimuthal dimension leads to an isotropic ($\cos\Theta, \Phi$) distribution,
as expected from the decay of a $J = 0$ particle.
Vice-versa and more interestingly, 
the average over $\cos\Theta$ leads to an isotropic dilepton decay distribution of the \jpsi.
On the other hand, it is now apparent that if the $\cos\Theta$ distribution is
not uniform or linear, so that $\langle \cos^2\Theta\rangle \ne 1/3$,
the dilepton distribution measured in the CC frame will not be isotropic
and the presence of a nonzero natural polarization $\lambda_0$ will be revealed.

\section{Kinematic relations}
\label{sec:kinematics}

The kinematic variables relevant for the following discussion are:
the B and \jpsi laboratory momenta, $\vec{P}$ and $\vec{p}$, respectively;
the \jpsi momentum in the B rest frame, $\vec{p}'$;
the B, \jpsi, and $X$ masses, $M$, $m$, and $m_X$, respectively; 
and $\cos\Theta$, where $\Theta$ is the \jpsi emission angle in the B HX frame.
In the following, we imply $\Theta \equiv \Theta_{\mathrm{HX}}$.

The $p_\perp$ and $p_\parallel$ \jpsi momentum components,
respectively perpendicular and parallel to the B momentum ($\vec{P}$) direction,
transform from the B rest frame to the laboratory frame 
according to the Lorentz boost defined by $\beta = P / \sqrt{M^2 + P^2}$,
\begin{equation}
p_\perp =  p_\perp' \, = \,  p' \sin\Theta   \, ,
\label{eq:p_perp}
\end{equation}
\vglue-4mm
\begin{equation}
p_\parallel = \frac{1}{\sqrt{1-\beta^2}} \; \left( \, p_\parallel' + \beta \sqrt{p'^2 + m^2} \, \right)
= \sqrt{1 + \frac{P^2}{M^2} } \,  p' \cos\Theta + \frac{P}{M} \sqrt{p'^2 + m^2} \, ,
\label{eq:p_parallel}
\end{equation}
where $P$ is the modulus of $\vec{P}$ and $p'$ is given by
\begin{equation}
p' \, = \, \frac{1}{2M} \sqrt{\big( \, M^2+m^2-m_X^2 \,\big)^2 - 4 M^2 m^2} \, ,
\label{eq:p_prime}
\end{equation}
which reduces to
\begin{equation}
p' \, \simeq \, \frac{M^2-m^2}{2 M}
\label{eq:p_prime_approx}
\end{equation}
when $m_X^2 \ll M^2 + m^2$,
a relation satisfied within $\sim$\,1\% for $X = \mathrm{K}$ in B decays.

To facilitate the illustration of the concept, 
we restrict our initial considerations to high-momentum measurements, 
where the B and \jpsi laboratory momenta are significantly larger than their masses,
\begin{equation}
P \gg M \quad \text{and} \quad p \gg m \, .
\label{eq:high-p-approx}
\end{equation}
This condition 
(only used for an easier illustration of the concepts, in this and the next sections)
is satisfied, for example, in most charmonium measurements at the LHC.
The relations shown in the remainder of this section are, therefore,
applicable quantitatively to decays of not-too-low \jpsi momentum.
Equations~\ref{eq:p_perp} and~\ref{eq:p_prime_approx}
imply the general inequality $p_\perp < (M-m) \, \sin\Theta$ and, therefore,
\begin{equation}
p_\perp \, \ll \, p \quad \mathrm{if} \quad p \, \gg \, M-m  \, .
\label{eq:negligible_p_perp}
\end{equation}
This means that $p \simeq p_\parallel$,
so that the vectors $\vec{p}$ and $\vec{P}$ can be considered to be parallel,
$\vec{p} \parallel \vec{P}$, and, as previously anticipated,
the CC frame becomes identical to the corresponding laboratory frame (e.g., the HX frame).

We can now quantify the effect of the approximation of Eq.~\ref{eq:high-p-approx}
on a polarization measurement,
considering that the angle $\delta_{\mathrm{CC}}$ between the two polarization axes 
is the angle between the vectors $\vec{p}$ and $\vec{P}$,
given by \mbox{$\sin \delta_{\mathrm{CC}} = p_\perp/p$}.
Assuming that the decay distribution is of the kind
$\propto 1 + \lth^{\mathrm{CC}} \cos^2 \vartheta$ in the HX-CC frame
(i.e., $\lth^{\mathrm{CC}} \equiv \lambda_0$),
the corresponding \lth value in the HX frame is
(Eq.~21 of Ref.~\cite{Faccioli:EPJC69}, setting \lph and \ltp to zero)
\begin{equation}
\lth^{\mathrm{HX}} \, = \, \lth^{\mathrm{CC}} 
\frac{1- 3/2 \, \sin^2 \delta_{\mathrm{CC}}  }{1 + 1/2 \, \lth^{\mathrm{CC}} \sin^2 \delta_{\mathrm{CC}}}
\simeq \, \left[ 1 - \frac{3 + \lth^{\mathrm{CC}}}{2} \sin^2 \delta_{\mathrm{CC}} \right] \, \lth^{\mathrm{CC}}  \, ,
\label{eq:CC_to_HX_lth}
\end{equation}
where the approximate equality is valid in the small angle limit. 
Therefore, the relative deviation of a \lth measurement in the HX frame from its CC expectation,
\begin{equation}
\left| \frac{\lth^{\mathrm{HX}}-\lth^{\mathrm{CC}}}{\lth^{\mathrm{CC}}} \right| \,
\simeq \, \frac{3 + \lth^{\mathrm{CC}}}{2} \sin^2 \delta_{\mathrm{CC}} \, 
\le \, \frac{3 + \lth^{\mathrm{CC}}}{2} \left(\frac{M-m}{p}\right)^2  \, ,
\label{eq:error_CC_to_HX_lth}
\end{equation}
is at most of order 2--4\%, depending on \lth,
for the polarization of \jpsi mesons of $\pt = 10$\,GeV and rapidity $y=1$, 
and decreases with increasing \pt and $|y|$.

We will now use the condition $\vec{p} \parallel \vec{P}$, mentioned above.
Taking $p_\parallel$ in Eq.~\ref{eq:p_parallel} as expression for $p$, with 
$$\sqrt{p'^2 + m^2} = \frac{M^2+m^2}{2 M} 
\quad \textrm{and} \quad
\sqrt{1 + \frac{P^2}{M^2} } \, \frac{p'}{P} \simeq \frac{p'}{M}\, ,$$
the first relation deriving from Eq.~\ref{eq:p_prime_approx}
and the second from the $P \gg M$ assumption,
we find that
\begin{equation}
\vec{p} \, \simeq \, \vec{P} \,\, f(\cos\Theta)  \, , 
\label{eq:p_vs_P_approx}
\end{equation}
with
\begin{equation}
f(\cos\Theta) = \left(  \frac{1- \cos\Theta}{2} \, \frac{m^2}{M^2}  \, + \frac{1+ \cos\Theta}{2}   \right) \, .
\label{eq:p_vs_P_approx_f}
\end{equation}
We conclude that the two vectors are related by a function that depends linearly on $\cos\Theta$.

This vector relation can be rewritten (remaining formally identical)
with $\vec{p}$ and $\vec{P}$ replaced by their moduli ($p$ and $P$)
or by their transverse (\pt and $P_\mathrm{T}$) 
or longitudinal ($p_\mathrm{L}$ and $P_\mathrm{L}$) components;
the formulas shown in the next section remain correct for all these options.
Even though the LHC experiments use \pt and $P_\mathrm{T}$
to present the polarization parameters 
(\lth, \lph, and \ltp, in several frames),
we will retain the simpler $p$ and $P$ notation.

\section{Effects of the experimental observation}
\label{sec:observation}

Typical polarization measurements usually introduce sculpting effects on the $\cos\Theta$ distribution.
If all \jpsi mesons produced in B decays were included in the analyzed data sample, 
the distribution would remain uniform, as it is at the production level.
In general, however, this is not possible in a real experiment,
and not only because of the selection criteria applied to improve the quality of the signal reconstruction:
the simple fact that we are \emph{observing} a sample of \jpsi mesons 
(rather than a sample of B mesons) 
and are limiting the range of their laboratory (transverse and/or longitudinal) momenta, 
is sufficient to distort the $\cos\Theta$ distribution.

Equations~\ref{eq:p_vs_P_approx} and~\ref{eq:p_vs_P_approx_f} can be read as follows.
If we consider a sample of events where the B meson is always produced with
a fixed laboratory momentum, of modulus (or component) $P$,
then the laboratory momentum (or component) $p$ of the \jpsi
is distributed uniformly between $(m/M)^2 \, P$ and $P$,
corresponding to the extremes $-1$ and $+1$ of the (natural) uniform distribution of $\cos\Theta$.

Having a sample of events distributed within a narrow interval around $P$ is not a realistic situation, 
given that the kind of experiment we are considering does not reconstruct the B kinematics 
or, anyway, does not perform the measurement as a function of $P$.
Instead, the experiment detects the \jpsi and, for each event, determines its momentum (or component) $p$,
so that the event sample is characterized by a distribution of $p$ (and not $P$) values.
However, while fixing $P$ leads to a uniform $\cos\Theta$ distribution, fixing $p$ does not.
In fact, for a narrow interval in $p$, neither the $\cos\Theta$ nor the $P$ distributions are uniform,
their ratio being
\begin{equation}
\frac{\mathrm{d} N}{\mathrm{d} \cos\Theta} \; \bigg/ \; \frac{\mathrm{d} N}{\mathrm{d} P} 
\equiv \, \left| \frac{\mathrm{d}P}{\mathrm{d} \cos\Theta} \right|
= \, \left\{
\begin{array}{l}
	\frac{1}{2} \, p        \, \left[ 1 - (m/M)^2 \right] \, f(\cos\Theta)^{-2}  \\[3mm]
	\frac{1}{2} \, p^{-1} \, \left[ 1 - (m/M)^2 \right] \, P^{2}
\end{array}
\right. \, ,
\label{eq:ratio_cosTheta_P_distr}
\end{equation}
where we made explicit the dependence on $\cos\Theta$ or $P$.

We see that, given a value of $p$,
the $\cos\Theta$ distribution would only be uniform
if the sample were chosen with a $\mathrm{d}N / \mathrm{d}P$ distribution proportional to $P^{-2}$.
Analogously, it would only be possible to obtain a uniform $P$ distribution
if $\cos\Theta$ would be distributed as $f(\cos\Theta)^{-2}$.
But, of course, we cannot chose the $P$ distribution,
which is precisely what physically determines the $p$ distribution of the event sample under analysis.

We conclude that, given a collected \jpsi sample,
how the (unobserved) $\cos\Theta$ distribution departs from a constant distribution
depends on the unknown shape of the unobserved $P$ distribution and, hence,
on the shape of the observed $p$ distribution.
To figure out how, we start by writing the ``original'' two-dimensional $(P, \cos\Theta)$ distribution as
\begin{equation}
\frac{\mathrm{d} N}{\mathrm{d} P \, \mathrm{d} \cos\Theta} \propto \left( \frac{M}{P} \right)^\rho \, ,
\label{eq:P_cosTheta_distr}
\end{equation}
where there is no $\cos\Theta$ dependence (constant distribution)
and the $P$ dependence is parame\-trized with a power-law function,
which is always a good approximation in a sufficiently narrow kinematic range.
In practice, Eq.~\ref{eq:P_cosTheta_distr} implies that 
in hypothetical measurements performed as a function of the B momentum, 
the $\cos \Theta$ distribution would remain flat 
(in the absence of other perturbing effects) 
and, hence, a fully smeared \jpsi polarization 
(negligible and $P$-independent $\lambda$ parameters)
should be seen.

We want to find the corresponding two-dimensional $(p, \cos\Theta)$ distribution,
which is the one relevant for the description of a measured data sample, 
where $p$, rather than $P$, is the available observable.
In the variable replacement
\begin{equation}
P \to p = P \, f(\cos\Theta) \, , \quad \cos\Theta \to \cos\Theta \, ,
\end{equation}
the measure changes as
\begin{equation}
\mathrm{d}P \; \mathrm{d} \cos\Theta = 1/f(\cos\Theta) \; \mathrm{d}p \; \mathrm{d} \cos\Theta \, ,
\end{equation}
so that Eq.~\ref{eq:P_cosTheta_distr} transforms into
\begin{equation}
\frac{\mathrm{d}N}{\mathrm{d}p \, \mathrm{d} \cos\Theta} \; \propto \;
\left( \frac{m}{p} \right)^\rho f(\cos\Theta)^{\,\rho-1} \, .
\label{eq:p_cosTheta_distr}
\end{equation}

\begin{figure}[t]
\centering
\resizebox{0.75\linewidth}{!}{\includegraphics{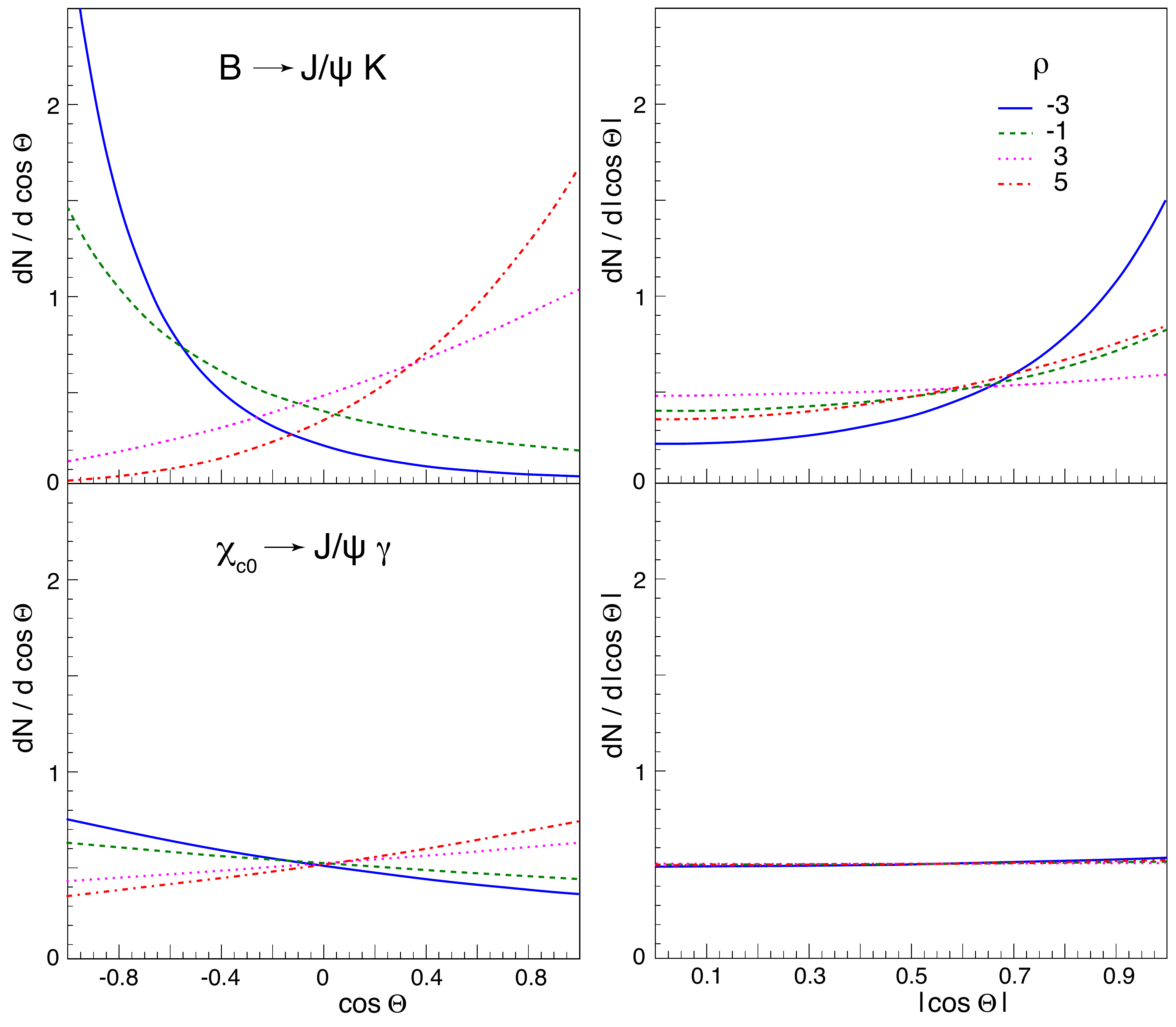}}
\caption[]{Examples of $\cos\Theta$ (left) and $|\cos\Theta\, |$ (right) distributions
for the decays $\mathrm{B} \to \jpsi \, \mathrm{K}$ (top)
and $\chi_{c0} \to \jpsi \, \gamma$ (bottom),
when the experiment selects samples of \jpsi mesons
having $p$ distributions $\propto ( p / m )^{-\rho}$, 
for several values of $\rho$.}
\label{fig:cosTheta_examples}
\end{figure}

Apart from finding an unchanged power-law dependence on momentum,
we see from this expression and from the $f(\cos\Theta)$ definition (Eq.~\ref{eq:p_vs_P_approx_f})
that the effective $\cos\Theta$ distribution will, in general, 
depart from the flat or linear shape that leads to the condition 
$\langle \cos^2\vartheta \rangle = 1/3$ (perfectly spherical distribution).

Figure~\ref{fig:cosTheta_examples}-left shows examples of $\cos\Theta$ distributions
for the $\mathrm{B} \to \jpsi \, \mathrm{K}$ decay (top) and, 
for illustration, for the $\chi_{c0} \to \jpsi \, \gamma$ decay (bottom),
for several values of the exponent $\rho$.
In their analytical description, these two decays only differ by the value of $m/M$,
which fully determines the shape of $f(\cos\Theta)$. 
The corresponding $|\cos\Theta\, |$ distributions,
shown in Fig.~\ref{fig:cosTheta_examples}-right,
are the ones relevant for the effect under study:
whether or not they are flat, and to what degree,
determines if the observed polarization, 
in the HX or CS frame approximating the CC frame,
will be fully smeared or only attenuated.
The cases $\rho=1$ and $\rho=2$ lead, for both decays,
to a flat $|\cos\Theta\, |$ distribution and are not shown 
in Fig.~\ref{fig:cosTheta_examples}.

The deviation from a flat distribution is larger for larger values of $| \, \rho-1 \, |$.
It is important to understand that $\rho$ is the (locally defined) ``slope'' of the $p$ distribution
of the collected \jpsi sample,
affected by experimental acceptance and efficiency effects.
Even if those effects are taken into account and corrected for in 
the measurement of the dilepton angular distribution,
and even if those corrections bring the $p$ distribution close to its natural shape,
the $\cos\Theta$ distribution, defined in the B rest frame
and additionally affected by the acceptance and reconstruction efficiency of $X$, 
is not observed and, hence, cannot be corrected.
It is, therefore, the raw experimental distribution of $p$, before corrections, 
that determines the sculpting of the $\cos\Theta$ distribution.

For example, the lepton selection criteria can induce 
a turn-down shape for the lowest detected values of $p$,
so that the $\rho$ exponent is negative at low $p$,
zero at the maximum of the distribution, and positive at high $p$.
Correspondingly, with varying $p$ the shape of the $\cos\Theta$ distribution will change,
according to Eq.~\ref{eq:p_vs_P_approx_f},
leading to varying degrees of smearing of the polarization observed in the CC 
(i.e., HX or CS) frame.
Effectively, therefore,
the smearing of the polarization is strongly influenced by purely experimental features of the measurement,
which may be difficult to account for.
This fact can lead to disagreements between experiments performing the same measurement
with different detectors and selection criteria, as illustrated in the next section.

Equation~\ref{eq:p_vs_P_approx_f} shows another factor that influences the strength of the smearing:
the dependence of $f$ on $\cos\Theta$ vanishes in the limit $m/M \to 1$.
We expect, for example, a significantly stronger polarization smearing for \jpsi mesons
from \mbox{$\chi_{c0} \to \jpsi \, \gamma$}
(where, incidentally, $\lambda_0 = +1$)
than from $\mathrm{B} \to \jpsi \, \mathrm{K}$,
given the similarity of the $\chi_{c0}$ and \jpsi masses.
This effect is clearly seen by comparing the top and bottom panels of Fig.~\ref{fig:cosTheta_examples}.

The above description is appropriate for measurements made at high momentum-to-mass ratio.
In particular, in the $\chi_{c0}$ decay the condition of Eq.~\ref{eq:negligible_p_perp},
leading to $\vec{p} \parallel \vec{P}$, 
is practically always satisfied and the ``ordinary'' \jpsi HX (or CS) axis,
adopted in the measurement,
becomes coincident with the corresponding CC axis over the entire momentum range of the measurement.
Moreover, for $\pt \gtrsim 10$\,GeV and even at midrapidity,
the analytical relation of Eq.~\ref{eq:p_vs_P_approx_f} is almost exact
and the previous discussion should faithfully reproduce the reality.
In the case of B decays, a slightly higher threshold in \pt and/or $|y|$ is necessary.
As a strong counterexample, we can consider the typical events produced
by decays of the much heavier Higgs boson:
the $\mathrm{H} \to \jpsi \, \gamma$ decay, e.g., strongly departs from the assumed approximations.
It remains true that a smearing of the natural \jpsi polarization 
(transverse, in this case) 
is expected,
because the direction of the $z$ axis of the \jpsi HX frame 
(the observation frame)
is certainly not fully correlated with the emission direction of the \jpsi in the Higgs rest frame
(natural polarization axis).
However, the sizeable \jpsi momentum in the Higgs rest frame,
$p' \simeq M/2 \simeq 62.5$\,GeV, must play an important role,
determining different observations depending on whether the laboratory momentum $p$ 
is much smaller or much larger than $p'$: a complex smearing pattern should be seen.

\section{Effects of the event selection requirements}
\label{sec:selections}

\begin{figure}[t]
\centering
\resizebox{0.5\linewidth}{!}{\includegraphics{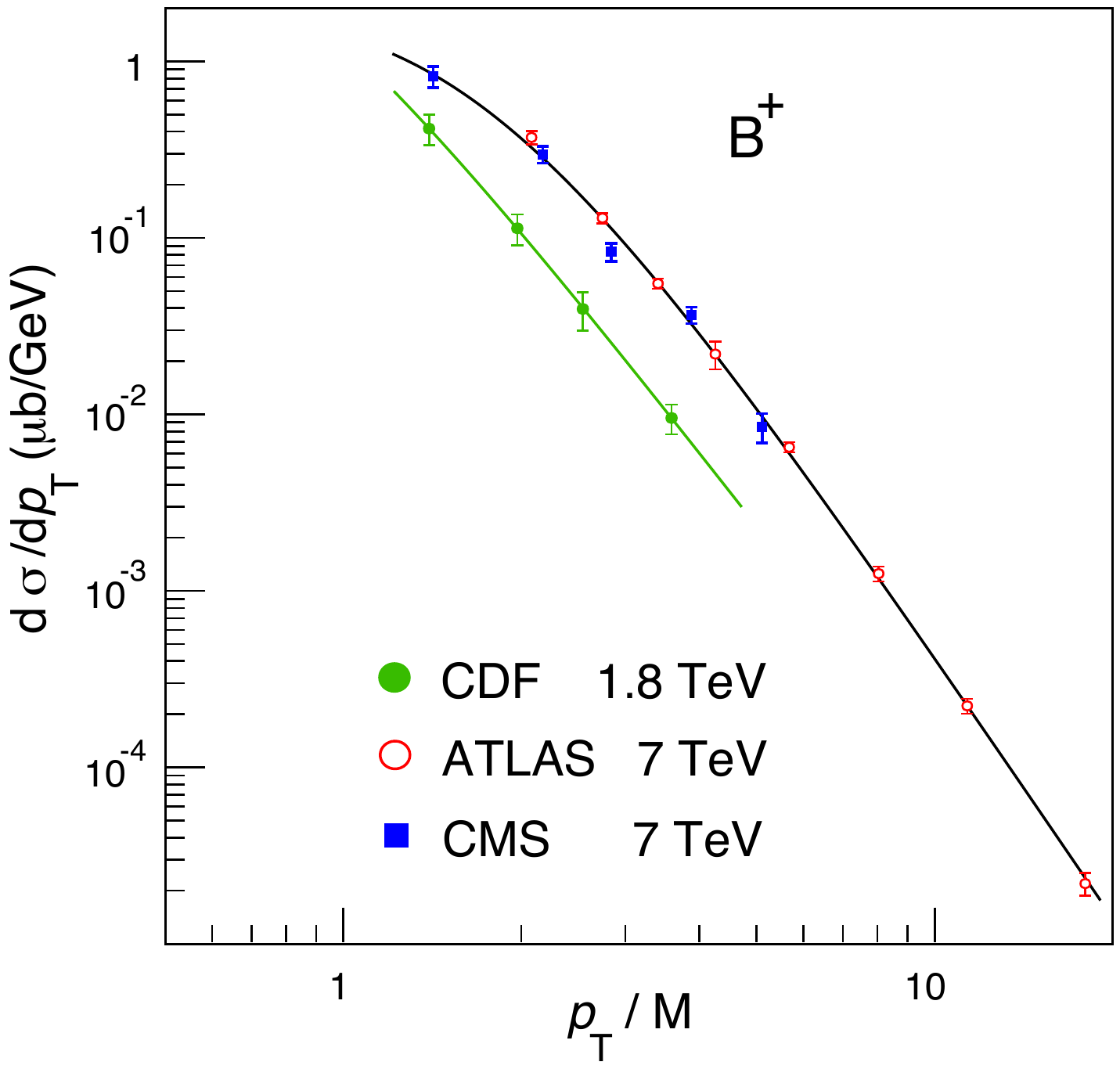}}
\caption[]{Midrapidity double-differential cross sections for the production of B$^+$ mesons 
at $\sqrt{s} = 1.8$ and 7~TeV, measured by
CDF~\cite{CDF:Bplus-xsection}, 
ATLAS~\cite{ATLAS:Bplus-xsection-7TeV}, and 
CMS~\cite{CMS:Bplus-xsection-7TeV}.
The curves represent empirical parametrizations
(Eq.~2 of Ref.~\cite{Faccioli:PLB773}).}
\label{fig:B_pT_distributions}
\end{figure}

In this section we describe, using simulated events, 
how (realistic) experimental selections of the analyzed sample 
can affect the measurement of the polarization of non-prompt \jpsi mesons.
As shown in the previous section, the observable polarizations depend, among other things, 
on the distribution of the \jpsi momentum component with respect to which the polarization is measured. 
In what follows, we refer to the transverse component, \pt.
The \jpsi kinematics are a direct reflection of those of the B meson. 
For the generation of the events, we have used a realistic \pt distribution, 
obtained by parametrizing the B meson differential cross sections 
measured by ATLAS~\cite{ATLAS:Bplus-xsection-7TeV} 
and CMS~\cite{CMS:Bplus-xsection-7TeV}
in pp collisions at $\sqrt{s} = 7$~TeV,
shown in Fig.~\ref{fig:B_pT_distributions} as a function of $\pt/M$. 
In the event generation, the B decays into $\jpsi + X$ isotropically 
and the \jpsi decays into $\mu^+ \mu^-$ according to the 
distribution of Eq.~\ref{eq:J0decay_cHX_distr} in the cHX frame.

No approximations are made in the generation of the decay distributions,
but some hypotheses were made in the modelling of the physical composition of the sample
with the purpose of simplifying the discussion of the effects that we want to emphasise.
First, we assume that $X$ is a $J=0$ particle (kaon, pion, $\eta$ meson, etc.),
so that the \jpsi mesons are necessarily produced with a well-defined natural polarization: 
longitudinal along the cHX axis.
In reality, other $\mathrm{B} \to \jpsi \, X$ decays exist, 
where $X$ is a vector particle or a multi-body system
(possibly having an invariant mass significantly larger than the relatively small mass of the kaon).
In such cases, without the $J(X) = 0$ constraint, 
the \jpsi mesons are no longer produced with a fully longitudinal natural polarization,
and our result becomes an upper limit for the \emph{magnitude} of the observed polarization.
Additionally, inclusive non-prompt production includes more complex decay chains,
e.g., where the B meson first decays into a $\chi_{c1}$, a $\chi_{c2}$ or a $\psi$(2S) meson,
which then decays into a \jpsi.
Also this kind of further complexity, leading to a reduction of the observed polarization,
will be neglected in our exposition.
We will revise these assumptions in the next section,
where we will adopt a more realistic description of the non-prompt sample.

\begin{figure}[t!]
\centering
\resizebox{0.47\linewidth}{!}{\includegraphics{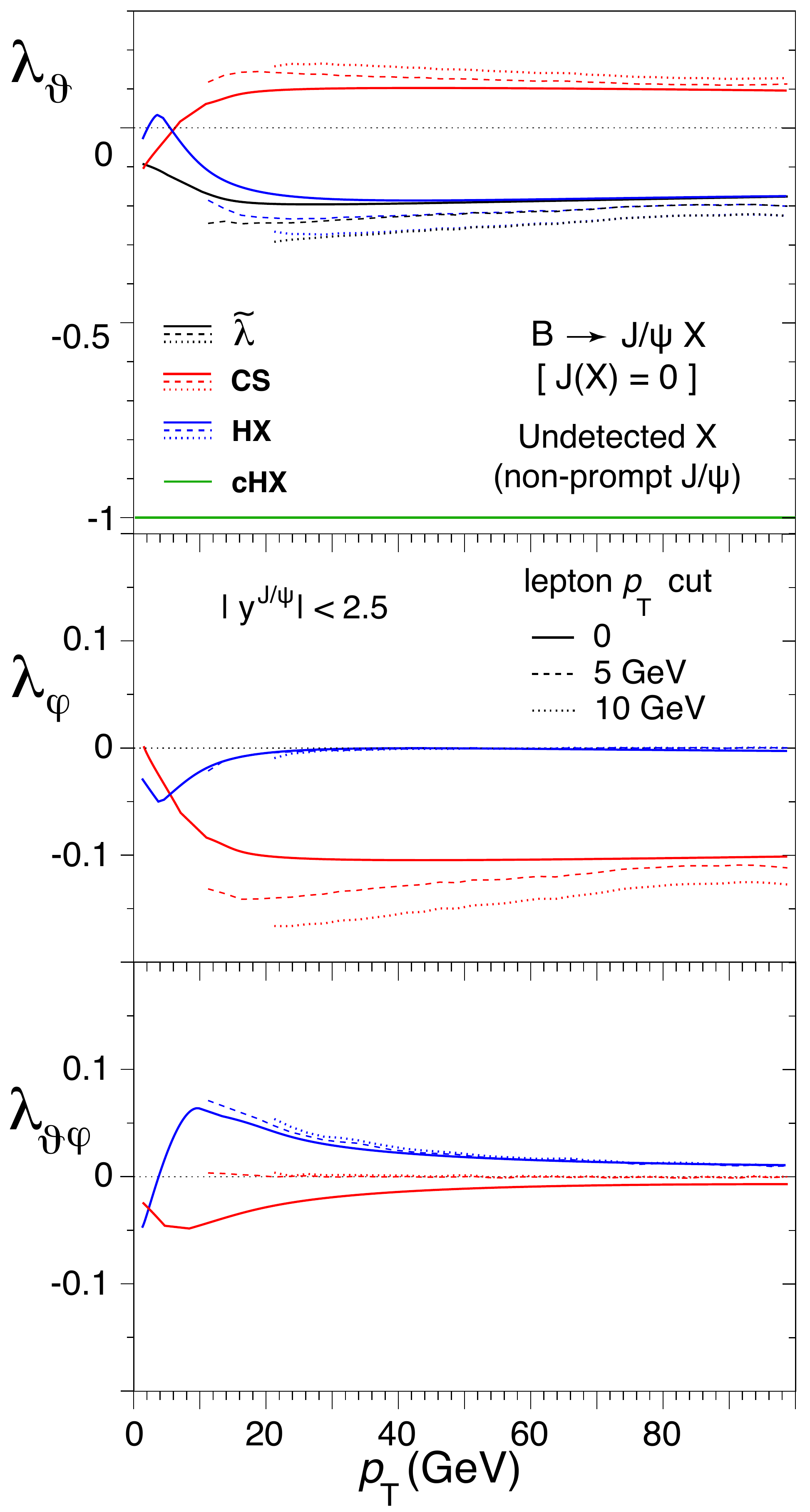}}
\caption[]{The frame-dependent anisotropy parameters \lth, \lph, and \ltp (top to bottom rows),
as well as the frame-invariant parameter $\ltilde$ (top row),
of the dilepton decay distribution of inclusively observed non-prompt \jpsi mesons.
The results are shown in the HX (blue) and CS (red) frames, 
as functions of the \jpsi \pt, with no further selection (solid) and 
when minimum \pt values are imposed on the decay muons (dashed and dotted).
The green line represents the natural polarization of the generated events, $\lth = -1$.}
\label{fig:B_to_JpsiK}
\end{figure}

As already mentioned, we consider a measurement where
a sample of \jpsi mesons from $\mathrm{B} \to \jpsi \, X$ decays 
is selected by requiring that the distance between the primary vertex (the pp collision) 
and the dimuon vertex (where the \jpsi decays)
is larger than a certain threshold value,
so as to reject all the promptly-produced \jpsi mesons.
The selection of this non-prompt \jpsi sample is made without using any information regarding
the accompanying particle $X$, which remains undetected.
The resulting polarization parameters of the \jpsi dilepton decay distribution,
\lth, \lph, and \ltp, in the HX and CS frames, as well as 
the $\ltilde$ frame-invariant parameter~\cite{Faccioli:lambdatildePRL,Faccioli:lambdatildePRD},
are shown in Fig.~\ref{fig:B_to_JpsiK};
a strong smearing of the natural longitudinal polarization ($\lth = -1$) is seen.
The residual polarization remains, nevertheless, quite pronounced and, furthermore,
shows a significantly non-flat \pt dependence.
The dashed and dotted lines illustrate the effect of performing the measurement 
using an event sample
where the decay muons are required to have minimum \pt values of 5 and 10\,GeV, respectively.
These selection criteria, inspired by thresholds applied in typical LHC analyses,
lead to non-negligible variations of the patterns.

\begin{figure}[p]
\centering
\resizebox{0.4\linewidth}{!}{\includegraphics{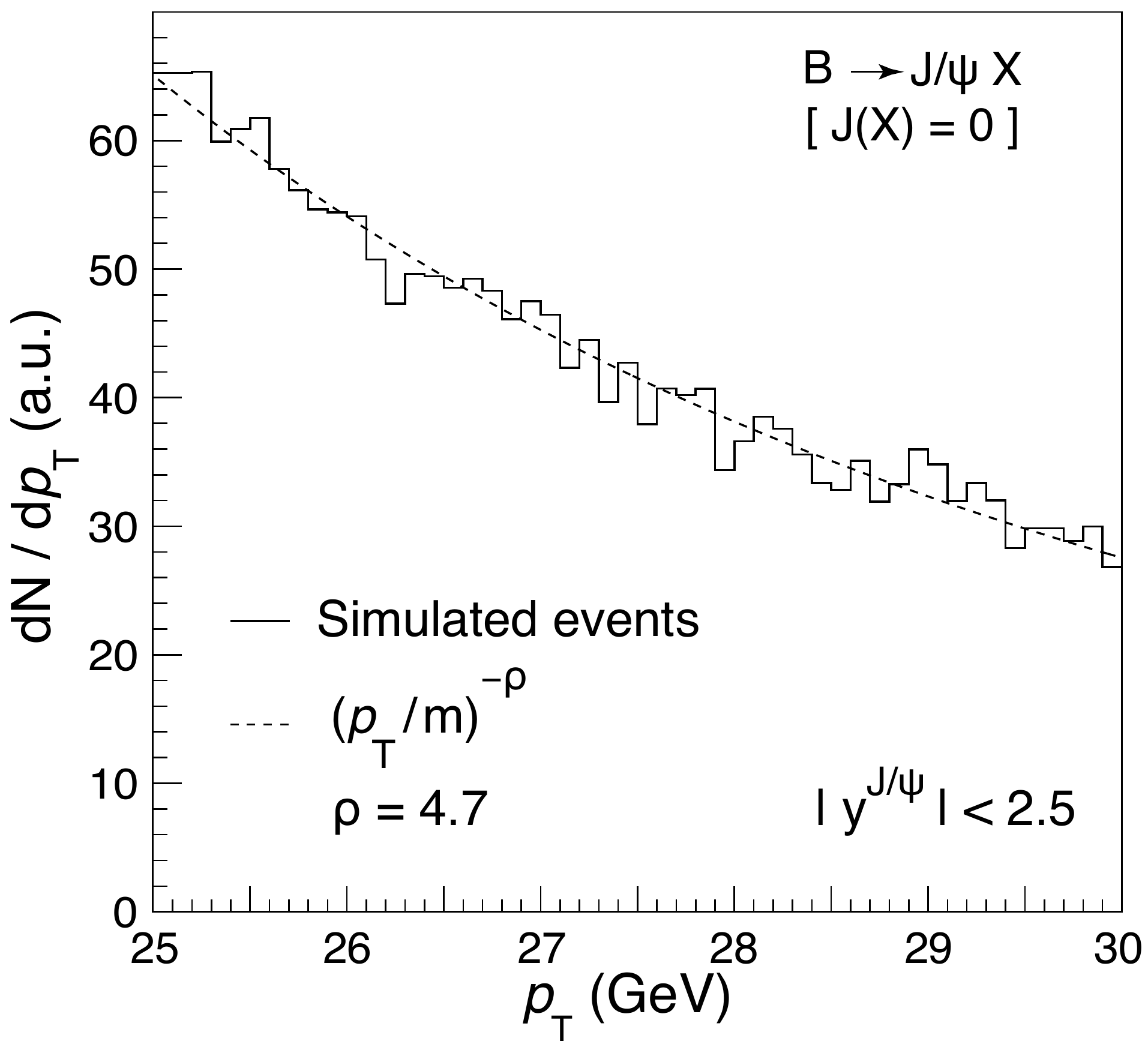}}
\caption[]{The \jpsi \pt distribution of simulated $\mathrm{B} \to \jpsi \, X$ events
with unobserved $X$, in the $25 < \pt^{\jpsi} < 30$\,GeV range.}
\label{fig:B_to_JpsiK_pT}
\vglue4mm
\centering
\resizebox{0.75\linewidth}{!}{\includegraphics{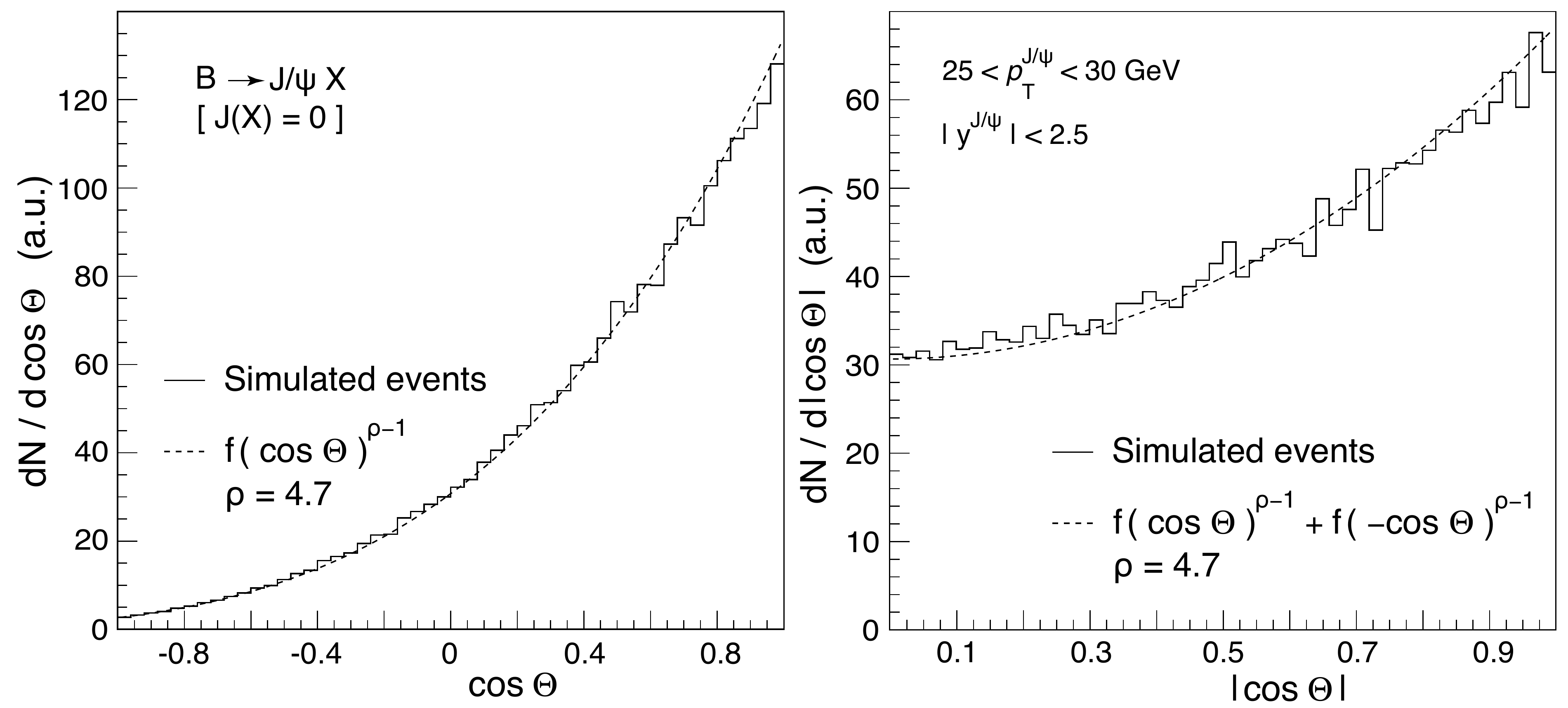}}
\caption[]{The $\cos \Theta$ (left) and $|\cos \Theta\, |$ (right) distributions,
in the B HX frame,
of simulated $\mathrm{B} \to \jpsi \, X$ events with unobserved $X$,
for $25 < \pt^{\jpsi} < 30$\,GeV.
The exponent $\rho$, determined by fitting the \pt distribution 
(Fig.~\ref{fig:B_to_JpsiK_pT}),
provides a good analytical description of the angular distributions.}
\label{fig:B_to_JpsiK_cosTH}
\vglue4mm
\centering
\resizebox{0.4\linewidth}{!}{\includegraphics{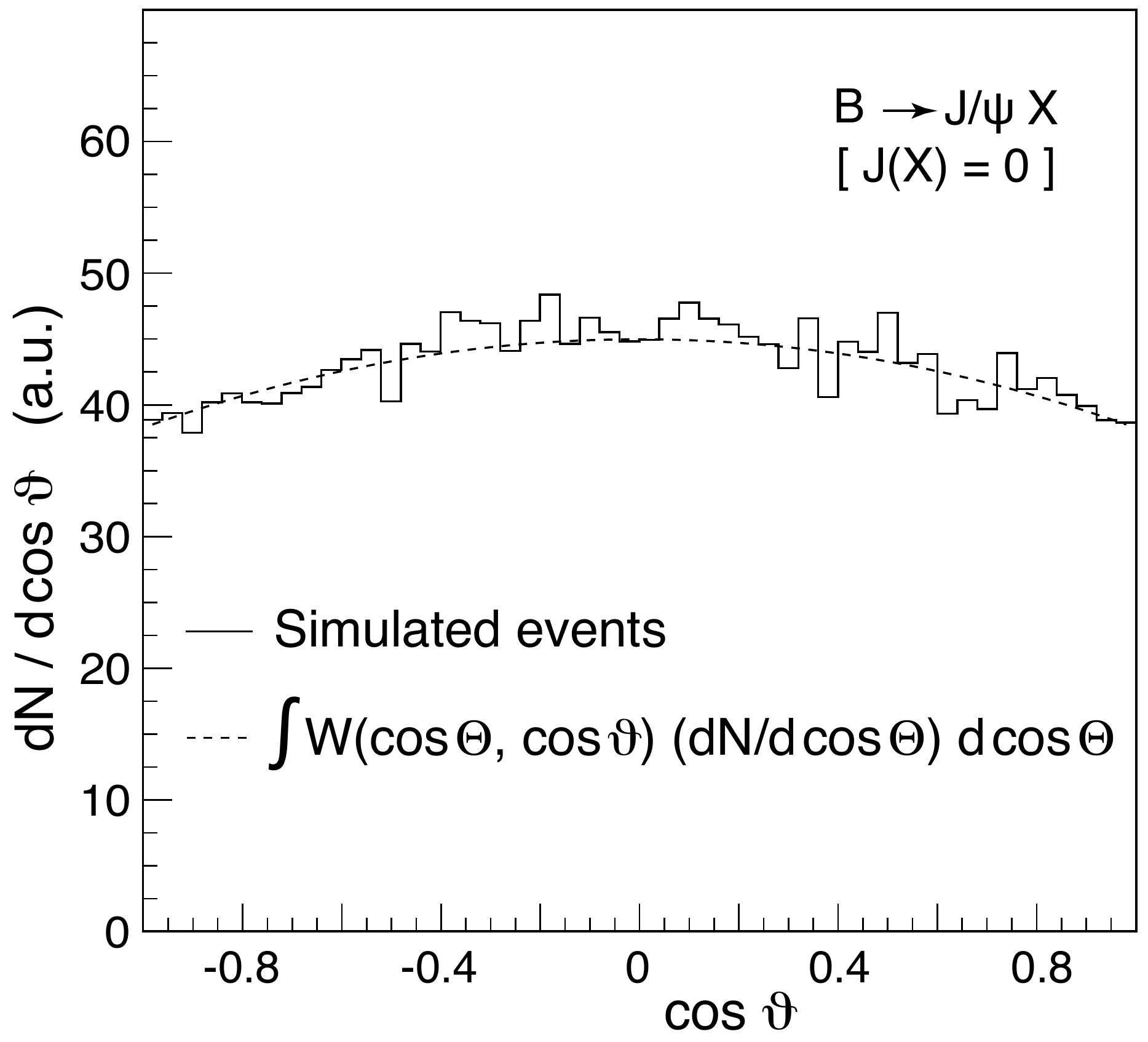}}
\caption[]{Distribution of $\cos \vartheta$, in the \jpsi HX frame,
corresponding to the one of $\cos \Theta$ shown in Fig.~\ref{fig:B_to_JpsiK_cosTH}.
The curve is obtained by integrating the full angular distribution $W$ 
(Eq.~\ref{eq:J0decay_CC_distr}) over $\cos \Theta$,
taking into account the modulation in the latter distribution.}
\label{fig:B_to_JpsiK_exclusive_costheta}
\end{figure}

The observed effects can be interpreted in the context of the analytical description 
presented in the previous section.
For this purpose we consider the 25--30\,GeV \jpsi \pt range,
where the high-momentum approximation, adopted in that discussion, is well satisfied.
Figure~\ref{fig:B_to_JpsiK_pT} shows the \pt distribution in the considered interval,
well reproduced by a decreasing power-law function 
with best-fit exponent \mbox{$\rho = 4.7 \pm 0.4$},
where the uncertainty is dominated by the size of the simulated event sample.

The central value of the exponent can be univocally converted
(Eqs.~\ref{eq:p_vs_P_approx_f} and~\ref{eq:p_cosTheta_distr}, 
as well as Fig.~\ref{fig:cosTheta_examples})
into the prediction of the $\cos \Theta$ distribution, $f(\cos \Theta)^{\,\rho-1}$,
meant to be measured in the HX frame (as in the previous section).
As shown in Fig.~\ref{fig:B_to_JpsiK_cosTH},
this prediction agrees well with the simulated distribution.
The $|\cos\Theta\, |$ distribution is not flat:
the smearing effect is only partial, 
justifying that \lth remains nonzero.
Figure~\ref{fig:B_to_JpsiK_exclusive_costheta} shows how 
the corresponding $\cos \vartheta$ distributions
can be reproduced by integrating the angular distribution $W$
(Eq.~\ref{eq:J0decay_CC_distr}) over $\cos \Theta$.
The integration would lead to a flat $\cos \vartheta$ distribution if
the $\cos \Theta$ distribution were flat (or linear).
The observed $\cos \vartheta$ modulation 
is a reflection of the non-uniformity of the $\cos \Theta$ distribution.
A slightly longitudinal polarization is observed.

\begin{figure*}[t!]
\centering
\resizebox{0.98\linewidth}{!}{\includegraphics{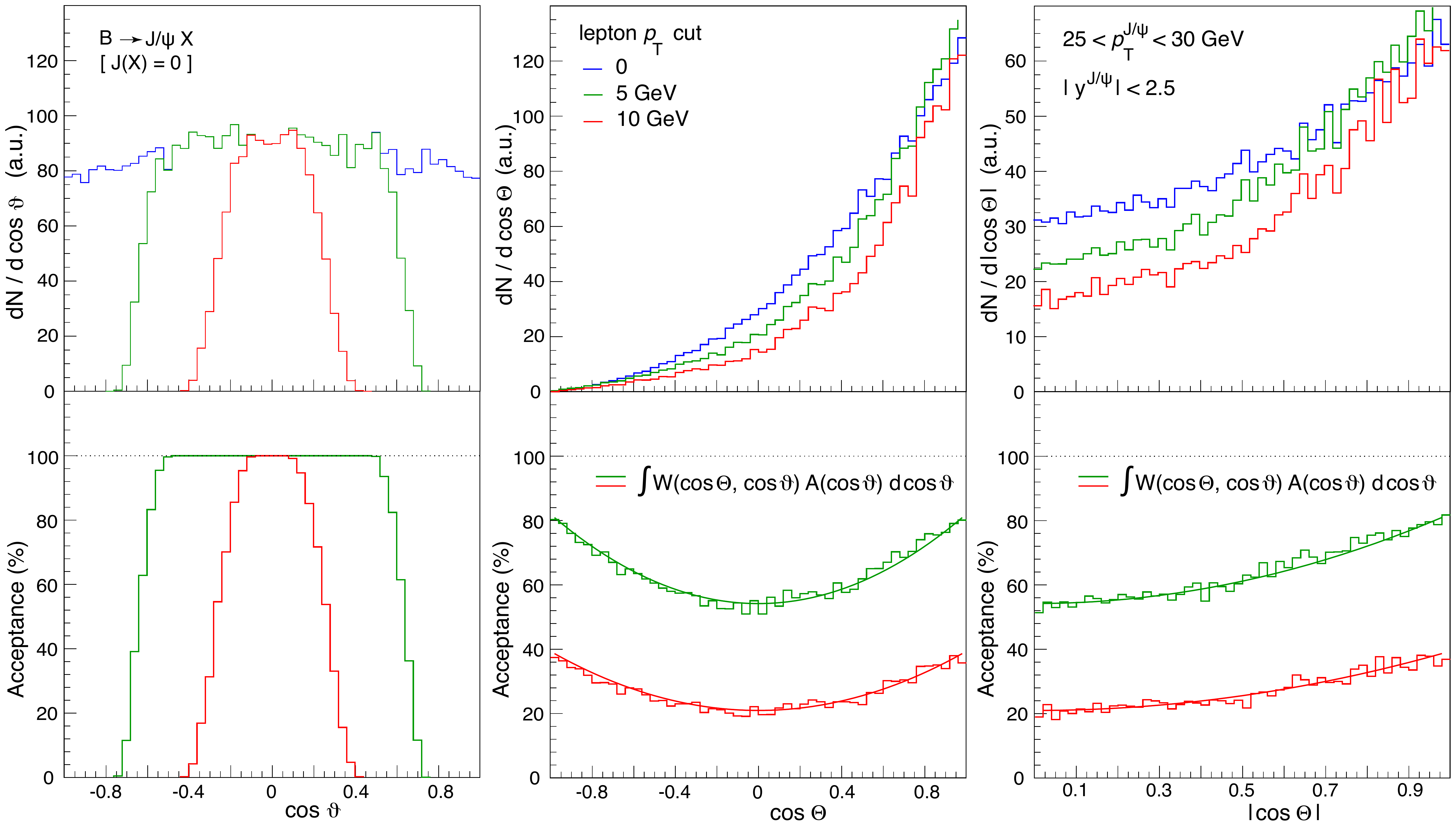}}
\caption[]{Top: 
The $\cos \vartheta$ (in the \jpsi HX frame, left), 
$\cos \Theta$ (in the B HX frame, middle), and
$|\cos \Theta\, |$ (right) distributions,
for simulated $\mathrm{B} \to \jpsi \, X$ events 
with unobserved $X$ and $25 < \pt^{\jpsi} < 30$\,GeV,
before (blue) and after (green and red) rejecting events with muon \pt smaller than 5 or 10\,GeV.
The distributions are arbitrarily normalized, to emphasize the shape differences.
Bottom: Corresponding acceptance ratios, 
representing the fractions of events surviving the two considered muon selections.
See the text for details on the superimposed curves.}
\label{fig:B_to_JpsiK_interpretation_mucuts}
\end{figure*}

Figure~\ref{fig:B_to_JpsiK_interpretation_mucuts}
illustrates the effects of requiring minimum \pt values on the muons used in the \jpsi reconstruction.
Such requirements strongly sculpt the dilepton distribution.
The left-top panel shows the $\cos \vartheta$ distribution, in the HX frame,
before and after applying selection cuts on the \pt of the muons;
the corresponding acceptance ratios,
representing the fraction of events that survive those selection cuts,
are presented in the left-bottom panel.

As usual in experimental measurements,
an accurate correction procedure must be applied for the recovery of the physical result.
The dashed and dotted lines in Fig.~\ref{fig:B_to_JpsiK} indicate the results that
an experiment would obtain \mbox{\emph{after}} such corrections have been applied
in the data-analysis procedure:
they represent the \emph{physical} polarization of the selected sample of \jpsi mesons and,
naturally, their values will always remain,
irrespectively of the strength of the applied selections,
within the physical domain of the polarization parameters.

Interestingly, however, we see that the obtained polarization result 
still reflects residual traces of how the sample was selected.
This also implies that two experiments applying different selection criteria
will obtain different \emph{physical} results.
We could even say, therefore,
that the polarization of \jpsi mesons from B decays,
at a given collision energy and in given kinematic conditions,
is not a well-defined, measurable, observable.
The experiment-dependent event selection criteria must be included in
an extended definition of the ``kinematic domain''.
This is, actually, a general feature of analyses where the polarization
of an indirectly-produced particle is studied ignoring the event-by-event correlations
between the mother's and daughter's decay angles.

Before continuing with the discussion of this problem,
we should explain that these ``corrected results''
were determined using Eq.~\ref{eq:J0decay_CC_distr_lambdas},
with the $(\cos \Theta,\Phi)$-dependent quantities replaced by average values
calculated for event sub-samples of \pt in each considered \pt bin.
The HX dilepton decay parameters are obtained 
from the distributions of the HX $\cos \Theta$ and $\Phi$ B decay angles
(analogous considerations apply to the CS frame).
This procedure assumes that the CC frame can be replaced by the ordinary HX frame,
an approximation valid in the high-momentum limit,
with associated uncertainty quantified by Eq.~\ref{eq:error_CC_to_HX_lth}.

To clarify why experiments applying different selection criteria
will obtain different physical results,
we need to study how the sculpting of the (\jpsi decay) 
$\cos \vartheta$ distribution affects the
(B decay) $\cos \Theta$ distribution.
The concept is the same as illustrated above for the ``inverse'' effect of how different
$\cos \Theta$ modulations determine different $\cos \vartheta$ distributions:
the two distributions are intimately correlated and any experiment-induced modification
of one will have an effect on the other.
We note that the dilepton decay angles appearing in Eq.~\ref{eq:J0decay_CC_distr}
can be calculated in the HX frame because,
in the considered high-momentum limit,
the variables $\cos \vartheta$ defined in the CC and HX frames
are effectively equivalent.

The middle-top panel of Fig.~\ref{fig:B_to_JpsiK_interpretation_mucuts} compares
the $\cos \Theta$ distributions obtained before and after muon selections.
They are arbitrarily normalized so that the effect on the shapes can be more easily seen:
removing low \pt muons induces a loss of events that is more pronounced as $\cos \Theta \to 0$,
and the higher is the threshold, the larger is the event loss.
The net result, symmetric in $\cos \Theta$, is best represented by the acceptance ratios,
shown in the middle-bottom panel.
To confirm that it is the sculpting of the observed dilepton $\cos \vartheta$ distribution
that causes this shaping of the unobserved $\cos \Theta$ distribution,
we use the acceptance ratios $A(\cos \vartheta)$
shown in Fig.~\ref{fig:B_to_JpsiK_interpretation_mucuts}-left
as weights in the integration of the four-dimensional angular distribution $W$
(Eq.~\ref{eq:J0decay_CC_distr}) over $\cos \vartheta$.
While we know that a full and uniform $\cos \vartheta$ coverage
would lead to a uniform $\cos \Theta$ distribution
(in the absence of all other effects mentioned above),
using the distribution of the actually accepted dimuon events, $A(\cos \vartheta)$,
to perform the average over $\cos \vartheta$ leads to the green and red curves
shown in the middle-bottom panel of Fig.~\ref{fig:B_to_JpsiK_interpretation_mucuts},
which reproduce perfectly well the shapes of the acceptances as functions of $\cos \Theta$.
We conclude that the $\cos \Theta$ modulations induced by the muon selections
are a direct reflection of the sculpting of the $\cos \vartheta$ distribution.

The right panels of Fig.~\ref{fig:B_to_JpsiK_interpretation_mucuts}, 
identical to the middle ones except for the replacement of $\cos\Theta$ 
by its absolute value, $|\cos \Theta\, |$,
show even more clearly how the muon selections 
accentuate the unevenness of the B decay angular distribution,
therefore decreasing the smearing effect,
so that the observed polarization remains more significantly nonzero
(as seen in Fig.~\ref{fig:B_to_JpsiK}).

To remove the dependence of the measurement outcome
on the event selections specifically applied by the experiment, 
it is conceptually possible to adopt a fully four-dimensional analysis approach,
taking into account the acceptance correlations between
the $(\cos\Theta, \Phi)$ and $(\cos\vartheta, \varphi)$ angular variables.
However, this is not possible, by definition, when $X$ is not observed
and the angular analysis is, hence, restricted to its dilepton ``projection'',
as in the kind of measurements we are considering, which necessarily become
dependent not only on the kinematic domain covered by the experiment 
but also on analysis-dependent event selections.

We conclude this section with a comment on the importance of the 
mother-daughter mass difference, which can be illustrated by comparing the 
non-prompt \jpsi and $\psi$(2S) cases.
With the decrease of the momentum of the charmonium in the B rest frame,
\mbox{$p' \approx (M_\mathrm{B}^2 - M_{\psi}^2\,) \, / \, (2\,M_\mathrm{B})$},
from $p' \simeq 1.7$\,GeV for the \jpsi to $\simeq$\,1.3\,GeV for the $\psi$(2S),
it can be shown that
the smearing increases significantly and the magnitude of the observed
$\psi$(2S) polarization parameters 
turns out to be only about half of that seen in the \jpsi case.

\section{Polarization of non-prompt \jpsi}
\label{sec:nonprompt}

In the previous section we developed our illustration of experimental effects 
by considering a specific subcategory of non-prompt \jpsi events 
contributing to the inclusive measurement, 
those due to two-body decays of the kind $\mathrm{B} \to \jpsi \, \mathrm{K}$,
with K possibly replaced by another (relatively light) $J=0$ particle. 
Naturally, non-prompt events also result from other B decay channels,
the importance of which is discussed in this section.

The different decay topologies contributing to non-prompt \jpsi production 
have an interesting correspondence with process categories usually considered 
in the theoretical modelling of quarkonium formation.  
In fact, in an exclusive two-body decay, such as
$\mathrm{B} \to \jpsi \, X$, with $X=\mathrm{K}$, $\pi$ or $\eta$,
it is reasonable to assume that the formation of the \jpsi bound state 
often happens through the colour-singlet mechanism, 
where the decay immediately produces a colour-neutral 
${}^3S_{\,1}^{[1]}$ \ccbar state~\cite{Beneke:1998ks,Beneke:1999gq}. 
If, instead, the \jpsi is produced through an intermediate coloured state
(with possibly different quantum numbers)
the soft gluons emitted in the $\ccbar \to \jpsi$ colour neutralization 
should then recombine with the spectator quark of the B meson
to form exactly ``the right'' accompanying particle $X$, such as a kaon;
the corresponding probability should be small.

Intermediate octet \ccbar states should, however, 
have an important role in the ``cocktail'' of decays 
yielding an inclusive sample of non-prompt events.
Octet processes, where soft gluon exchanges happen, 
generally lead to multi-body final states,
where $X$ is a system of two or more particles.
Complex final states are also produced by decay chains, of the kind 
$\mathrm{B} \to Q \, X$, with the quarkonium 
$Q \equiv \chi_{c1}$ or $\chi_{c2}$ or $\psi$(2S)
subsequently decaying to a \jpsi.

From the point of view of the resulting observable \jpsi polarization, 
as it can be measured in an inclusive analysis,
the processes described above can be grouped in two categories:
a)~two-body B decays where $X$ is a kaon (or another relatively light particle)
and $J(X)=0$ (or, more generally, the \jpsi has longitudinal polarization);
and b)~all the remaining $\mathrm{B} \to \jpsi$ decays,
including multi-body configurations, two-body decays with $J(X) \ne 0$, and cascade sequences.
We will denote these two categories as \emph{two-body} and \emph{multi-body}.

The two categories differ as follows.
1)~The \jpsi mesons produced in two-body decays have
maximally longitudinal natural polarization ($J_{z'}=0$ in the cHX frame)
while those produced in the multi-body category have 
a significantly reduced polarization magnitude, 
reflecting a mixture of several decays with, in general, 
many kinds of $J_{z'}$ projections.
2)~For the two-body case, the hypothesis that $X$ is a kaon 
or any other relatively light particle determines
the value of the \jpsi momentum $p'$ in the B rest frame,
which is one of the parameters determining 
how the natural polarization is smeared when observed,
for example, in the \jpsi HX frame 
(Eq.~\ref{eq:p_prime_approx} shows that $p' \simeq 1.7$\,GeV);
instead, in multi-body decays 
the invariant mass $m_X$ of the accompanying system is a continuous distribution, 
including values significantly larger than the mass of a kaon, leading to smaller $p'$.
The smaller mother-daughter mass difference leads to a stronger smearing and,
hence, a smaller observable polarization.

\begin{figure}[t]
\centering
\resizebox{0.55\linewidth}{!}{\includegraphics{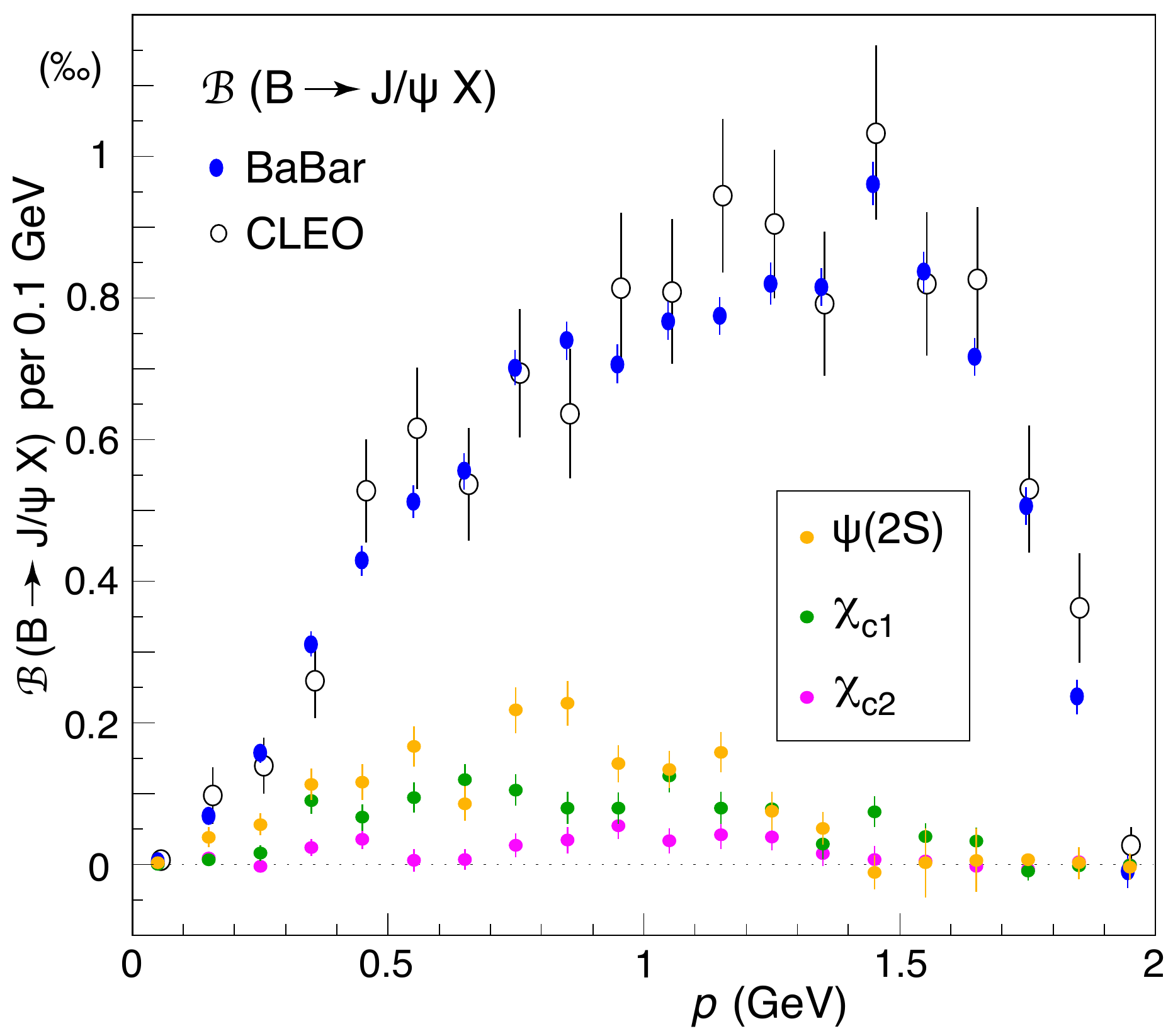}}
\caption[]{Laboratory momentum distribution of \jpsi mesons
emitted in decays of B mesons that are themselves produced by $\Upsilon$(4S) decays.
Besides the inclusive spectrum,
the figure also shows the individual contributions due to the feed-down decays
$\mathrm{B} \to \chi_{c1,2} \, X$ and $\mathrm{B} \to \psi\mathrm{(2S)} \, X$.}
\label{fig:Jpsi_Y4S_momentum_data}
\end{figure}

To quantify the importance of these different aspects,
we will consider the momentum distribution of \jpsi mesons emitted 
in the decays of $\mathrm{B}^+$ and $\mathrm{B}^0$ mesons produced almost at rest, 
in the decay of the $\Upsilon$(4S) resonance,
as reported by the CLEO~\cite{CLEO} and BaBar~\cite{BaBar} experiments.
To interpret those measurements, we must keep in mind that the B mesons 
have a momentum of only 0.33\,GeV in the $\Upsilon$(4S) rest frame and, 
therefore, the \jpsi momentum ($p$) distribution measured in the $\Upsilon$(4S) ``laboratory''
is a slightly smeared version of the one observed in the B rest frame ($p'$). 
The maximum shift produced by this smearing effect, 
calculated using a numerical simulation of the B meson production and decay process, 
is of order $p-p' \simeq 0.2$\,GeV.
For $m_X$ comparable to or lighter than the kaon mass, 
the value of $p'$ and, thus, the distribution of $p$ values
are practically independent of $m_X$ (Eq.~\ref{eq:p_prime_approx}). 
We can assume, therefore, that the ensemble of two-body decays, 
with $m_X \lesssim 0.5$\,GeV, 
is responsible for events in the range $p \gtrsim 1.7 - 0.2 = 1.5$\,GeV.
A large part of the measured $p$ distribution,
seen in Fig.~\ref{fig:Jpsi_Y4S_momentum_data},
covers a domain complementary to this,
showing the important role of multi-body decays.
For example, the decay chains $\mathrm{B} \to \chi_{c1} \to \jpsi$,
$\mathrm{B} \to \chi_{c2} \to \jpsi$, and $\mathrm{B} \to \psi\mathrm{(2S)} \to \jpsi$,
individually reported by BaBar and also shown in Fig.~\ref{fig:Jpsi_Y4S_momentum_data},
contribute mostly to the region $p < 1.5$\,GeV.

\begin{figure}[t]
\centering
\resizebox{0.6\linewidth}{!}{\includegraphics{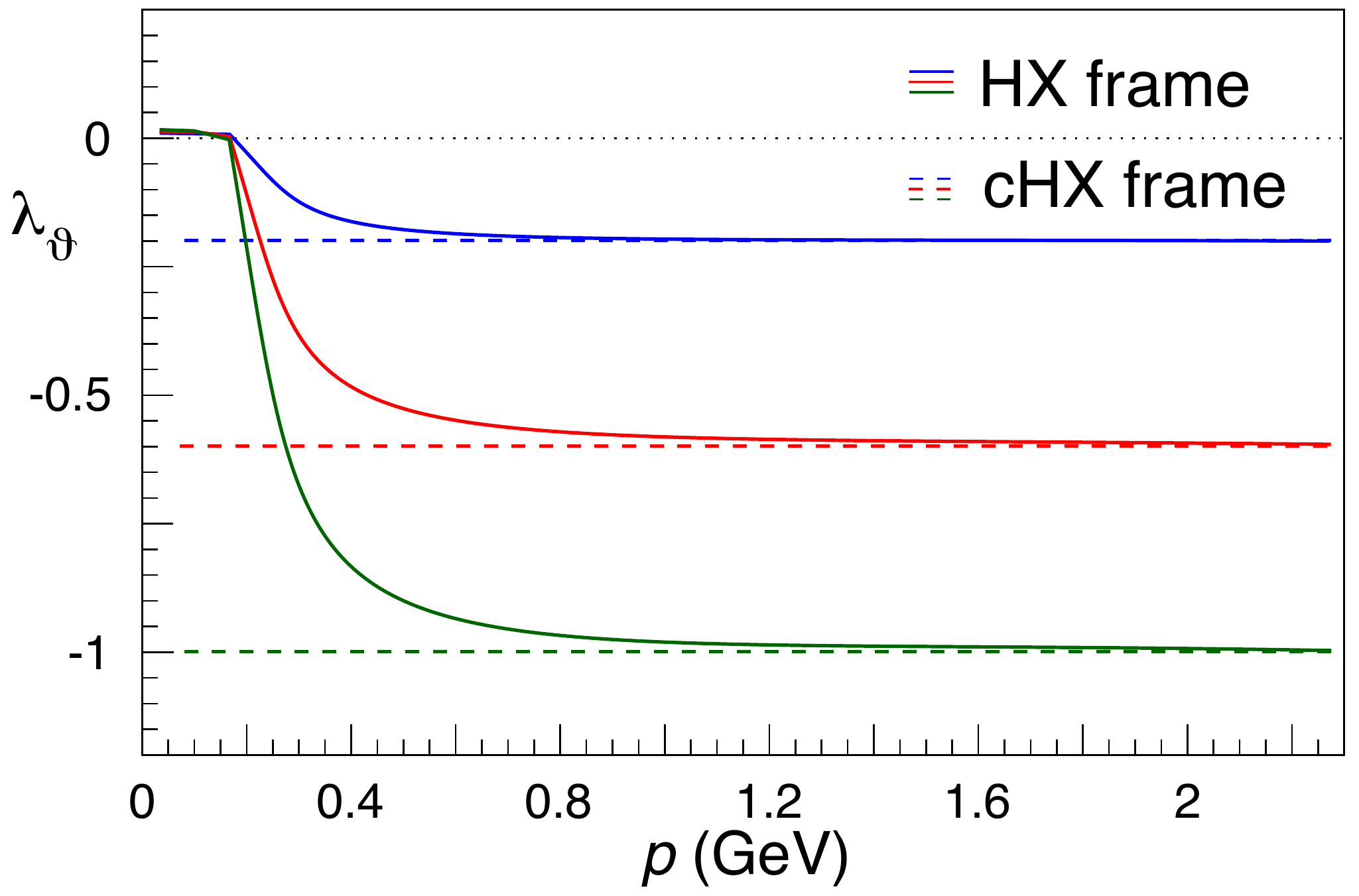}}
\caption[]{Polarization smearing from the cHX frame to the HX frame, 
in the $\mathrm{B} \to \jpsi \, X$ decay, for three different natural polarizations:
$\lth^{\mathrm{cHX}} \equiv \lambda_0 = -1$ (green), $-0.6$ (red), and $-0.2$ (blue).}
\label{fig:Jpsi_Y4S_lth_vs_p}
\end{figure}

In these experimental conditions, 
clearly very different from those of LHC measurements,
it also happens that the polarization measured in the HX frame, 
i.e., taking the direction of $\vec{p}$ as polarization axis,
will tend to be very close to the one measured in the cHX frame,
with polarization axis along $\vec{p'}$, given the similarity of the two momenta.
Figure~\ref{fig:Jpsi_Y4S_lth_vs_p} shows how \lth is smeared in the HX frame
with respect to the hypothetical natural polarization cases
$\lth^{\mathrm{cHX}} \equiv \lambda_0 = -1$, $-0.6$, and $-0.2$.
The first case corresponds to our hypothesis for the two-body processes:
the full longitudinal polarization for $p > 1.5$\,GeV remains practically unsmeared.
BaBar reported the values $\lth^{\mathrm{HX}} = -0.196 \pm 0.044$ for $p < 1.1$\,GeV
and $-0.592 \pm 0.032$ for $p > 1.1$\,GeV.
The value in the low-$p$ region refers to multi-body configurations.
We see that the polarization smearing for $\lth^{\mathrm{cHX}} \simeq -0.2$ 
leads to a $\lth^{\mathrm{HX}}-\lth^{\mathrm{cHX}}$ difference of order 0.01, 
as a weighted average over the events having $p < 1.1$\,GeV.
We will, therefore, assume the range from $-0.25$ to $-0.15$ for the average natural polarization
of the \jpsi mesons produced in multi-body decays.

As a cross check, we can try to interpret the result in the high-$p$ region,
which reflects a mixture of two-body and multi-body events.
Assuming, for simplicity,
that all the events in the range $p > 1.5$\,GeV are due to two-body processes,
we derive that these processes contribute
$(40 \pm 1)\%$ of the events in the $p > 1.1$\,GeV region.
Assuming $\lth^{\mathrm{cHX}} = \lth^{\mathrm{HX}}= -0.592 \pm 0.032$
for the average natural polarization of the mixture in the broader range
(the polarization smearing is practically nonexistent for $p > 1$\,GeV,
as seen in Fig.~\ref{fig:Jpsi_Y4S_lth_vs_p})
and taking $\lth^{\mathrm{cHX}}$ between $-0.25$ and $-0.15$ for the subsample of multi-body decays,
the sum rule in Eq.~32 of Ref.~\cite{Faccioli:EPJC69}, inverted,
leads to a value between $-1.1$ and $-0.9$ for the two-body polarization,
which is in perfect agreement with our assumption that this category of processes
leads to fully longitudinal \jpsi mesons.

We will now convert this information, derived from measurements made at the $\Upsilon$(4S) resonance,
into realistic expectations for the non-prompt \jpsi polarization 
as measurable in a high-energy collider experiment.
Here the unobserved B meson, generally produced with a large laboratory momentum,
emits the \jpsi almost collinearly (Eq.~\ref{eq:negligible_p_perp}),
so that the HX axis adopted for the observation of the dilepton decay loses its correlation to the natural (cHX) one,
and a significantly smeared polarization is observed, as we saw in Section~\ref{sec:selections}.

Considering the spectra measured by BaBar and CLEO,
and assuming that the transition to multi-body events happens below $p \simeq 1.5$\,GeV,
the fraction of two-body events can be quantified as $f_{2\mathrm{body}} = (22 \pm 1)\%$.
However, the relative contribution of two- and multi-body processes in (high-energy) hadron collisions
is probably not the same as the one observed in the conditions of BaBar and CLEO,
given that a different admixture of parent hadron species containing $b$ quarks
(additionally including $\mathrm{B}_s$ mesons and $b$ baryons)
contributes to the non-prompt \jpsi sample.
It is also conceivable that the proportion between the two kinds of topologies 
may be affected by experimental selections tending to remove events 
of one or the other category, a possibility that should be carefully studied during the analysis. 
For these reasons, besides the realistic mixture using $f_{2\mathrm{body}}$, 
we will also report the two separate predictions for the two-body and multi-body cases.
The measurement itself should be able to consider the two physical options
and determine their effective relative contributions.

The two-body expectation is obtained, as in Section~\ref{sec:selections},
assuming $\lth^{\mathrm{cHX}} = -1$ and $m_X = m_K = 0.5$\,GeV.
For the multi-body case, 
from the $p$ distribution of Fig.~\ref{fig:Jpsi_Y4S_momentum_data}
and the $m_X$-to-$p$ correlation described by Eq.~\ref{eq:p_prime}, 
assuming $p \simeq p'$,
we deduce that the spectrum of physical possibilities is reasonably well covered 
by $m_X$ values in the 1--2\,GeV range.
Taking into account that a higher $m_X$ value leads to a more strongly smeared polarization
in the experimental frames,
we can define reasonable upper and lower margins for the observable polarization magnitude:
they correspond, respectively, to the pairs of parameter values
$\lth^{\mathrm{cHX}} = -0.15$, $\langle m_X \rangle = 2$\,GeV, and
$\lth^{\mathrm{cHX}} = -0.25$, $\langle m_X \rangle = 1$\,GeV.

\begin{figure}[t]
\centering
\resizebox{0.5\linewidth}{!}{\includegraphics{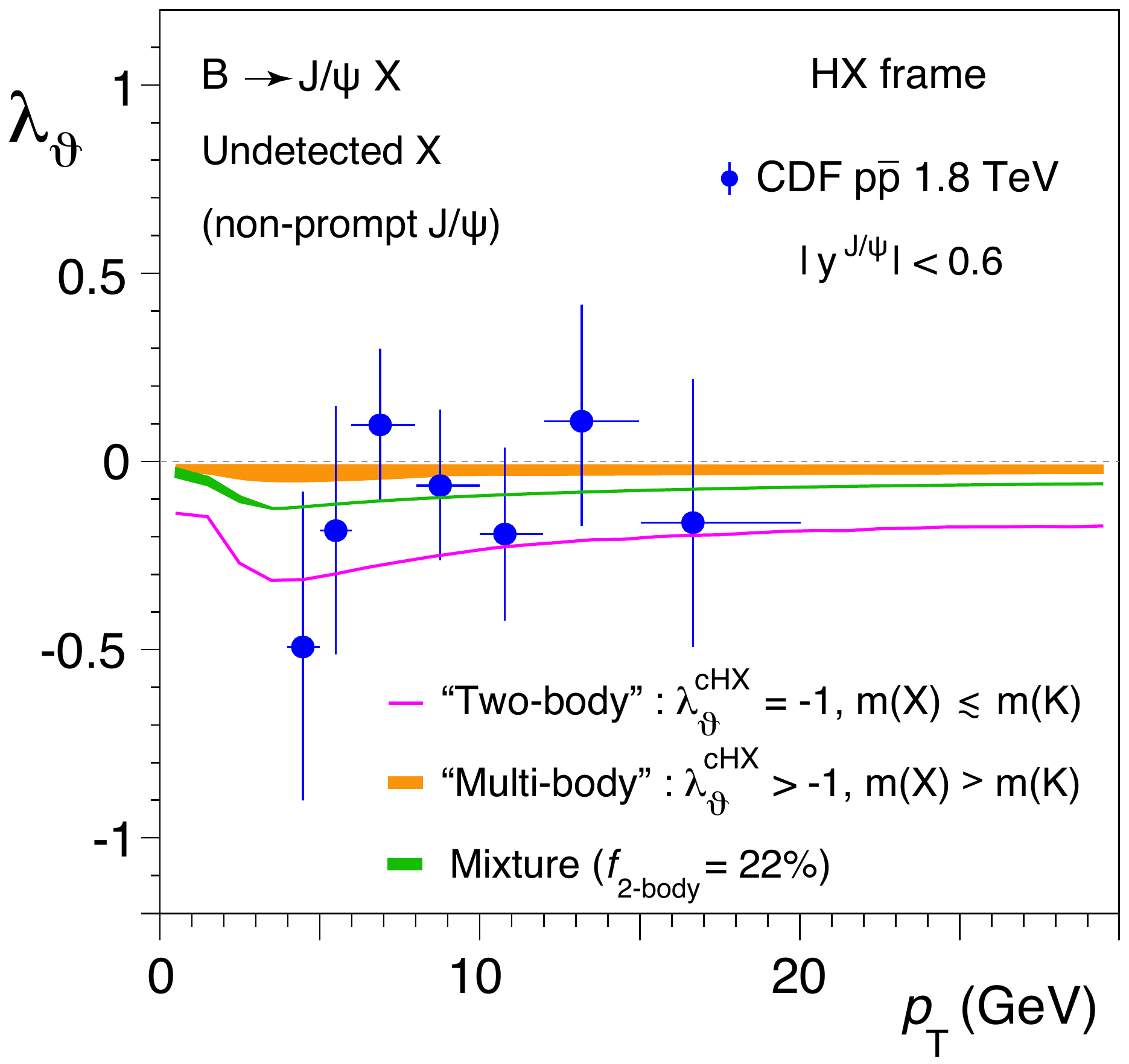}}
\caption[]{The non-prompt \jpsi polarization (\lth in the HX frame) measured by CDF~\cite{CDF:nonprompt}
in $\mathrm{p}\overline{\mathrm{p}}$ collisions at $\sqrt{s} = 1.8$~TeV, as a function of \pt,
in the rapidity range $|y| < 0.6$,
compared to predictions assuming that the \jpsi is produced
in \emph{two-body} (pink curve) or \emph{multi-body} (orange band) B decays,
the width of the band reflecting reasonable variations of the
$\lth^{\mathrm{cHX}}$ and $\langle m_X \rangle$ parameters.
The green band represents a mixture of these two kinds of processes, 
as motivated in the text.
The simulation used the B \pt distribution measured by CDF
(Fig.~\ref{fig:B_pT_distributions}).}
\label{fig:nonprompt_Jpsi_CDF}
\end{figure}

The only existing measurement in the case of hadron collisions was performed by CDF~\cite{CDF:nonprompt},
which reported \lth in the HX frame as a function of \pt.
The data are shown in Fig.~\ref{fig:nonprompt_Jpsi_CDF},
together with predictions reflecting the specific conditions of the experiment ($|y| < 0.6$);
curves representing lepton selection effects (as those shown in Fig.~\ref{fig:B_to_JpsiK})
are not included given that those effects are negligible with respect to the (rather large) experimental uncertainties.
The precision of the data is not sufficient to indicate if one or the other mechanism is predominant. 
The intermediate prediction assumes the same mixture of processes as in the $\Upsilon$(4S) measurements.
The multi-body prediction is compatible with (octet-dominated) 
NRQCD calculations of non-prompt \jpsi polarization~\cite{Fleming:1996,Krey:2002}.

\begin{figure}[t!]
\centering
\resizebox{0.8\linewidth}{!}{\includegraphics{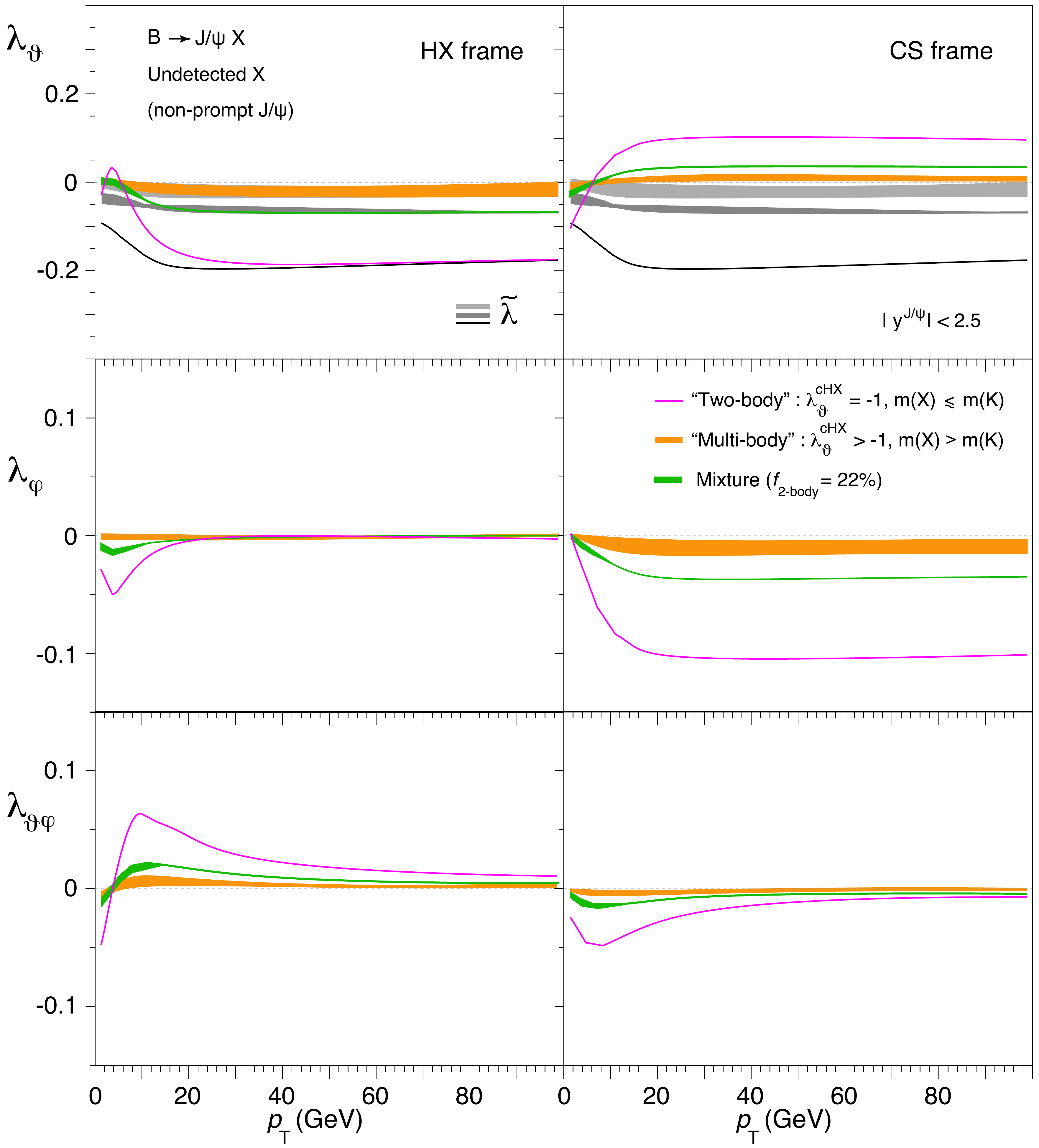}}
\caption[]{The frame-dependent anisotropy parameters \lth, \lph, and \ltp (top to bottom rows),
as well as the frame-invariant parameter $\ltilde$ (top row),
of the dilepton decay distribution of inclusively observed non-prompt \jpsi mesons,
in the two complementary assumptions that the \jpsi is produced 
in two-body B decays (pink curves, same as in Fig.~\ref{fig:B_to_JpsiK}) 
or in multi-body B decays (orange bands, 
their widths representing the variation of the relevant input parameters,
$\lth^{\mathrm{cHX}}$ and $\langle m_X \rangle$).
The green bands represent a mixture of these two kinds of processes, as motivated in the text.
The results are shown in the HX (left) and CS (right) frames, as functions of the \jpsi \pt.}
\label{fig:nonprompt_Jpsi}
\end{figure}

Figure~\ref{fig:nonprompt_Jpsi} shows predictions of all anisotropy parameters, 
both in the HX and CS frames,
calculated for the same ``typical LHC conditions'' that we considered in Section~\ref{sec:selections}.
In fact, the two-body curves were already shown in the left panels of Fig.~\ref{fig:B_to_JpsiK} and,
for visibility reasons, those reflecting the lepton selections are not repeated here.
Strictly speaking, these predictions are valid for $\sqrt{s} = 7$~TeV, 
the only energy-dependent ingredient being the shape of the B meson \pt-differential cross section. 
At significantly higher collision energies, a smaller polarization magnitude is foreseen in the high-\pt region, 
as a consequence of the expected difference in the asymptotic slope of the \pt distribution: 
as seen in Fig.~\ref{fig:cosTheta_examples} (comparing the $\rho = 3$ and~5 curves), 
a less steep \pt distribution leads to a flatter $|\cos \Theta\, |$ distribution and, 
hence, to an observable dilepton distribution closer to the full-smearing limit.

While the \jpsi mesons produced in two-body decays show sizeable polarizations 
($\lambda$~parameters significantly nonzero),
those produced in multi-body decays look practically unpolarized
(justifying the absence of multi-body curves representing the lepton selections,
which would be almost undistinguishable).
The almost complete lack of polarization also means that the multi-body prediction is essentially
insensitive to the assumptions made in our calculations:
reasonable variations in the input parameters $\lth^{\mathrm{cHX}}$ and $\langle m_X \rangle$
will not change the conclusion that multi-body decays lead to a barely detectable residual polarization.

For \jpsi $\pt \gtrsim 15$\,GeV, 
the difference between the multi-body and two-body polarizations is quite large, 
$\ltilde(\mathrm{multi{\text{-}}body}) - \ltilde(\mathrm{two{\text{-}}body}) \simeq 0.17$, 
with respect to the precision of measurements to be made 
using data collected during the Run~2 of the LHC.
We can conclude that such measurements should be able 
to determine the relative importance of these two kinds of processes.

\section{Discussion and summary}
\label{sec:summary}

We have studied how the polarization of non-prompt \jpsi mesons, 
indirectly produced in $\mathrm{B} \to \jpsi \, X$ decays, 
is observed in the typical conditions of inclusive quarkonium analyses in LHC experiments. 
In particular, we have derived the \pt dependences of the \lth, \lph, and \ltp anisotropy parameters 
of the dilepton decay angular distribution.

Non-prompt \jpsi mesons represent the main background for the study of 
prompt \jpsi production in high-energy colliders. 
Prompt \jpsi polarization measurements, crucial tests of the production models, 
can certainly benefit from the knowledge of the polarization of the non-prompt background, 
which, however, has not yet been measured at the LHC. 
One LHC experiment even reported the combined polarization 
of prompt and non-prompt \jpsi mesons~\cite{ALICE:Jpsi-pol-7TeV,ALICE:Jpsi-pol-8TeV}, 
without attempting a separation of the two samples, 
thereby reducing the impact of the measurement in its comparison to theory and other experimental results.
It is, therefore, interesting to study how the polarization of non-prompt \jpsi mesons 
would be measured by an experiment applying the same analysis technique 
to the prompt and non-prompt samples, 
except for the requirement of a minimum separation between the dilepton and primary vertices, 
which defines the non-prompt sample.

It is important to remark that such a polarization measurement, 
not involving the reconstruction of the B meson 
and referring to axes defined on the basis of the directions of the colliding protons, 
is conceptually and effectively very different from a measurement of 
the ``natural'' polarization of the \jpsi mesons produced in the B decays.
The natural polarization can be measured using the B meson direction as reference; 
however, this implies that the B meson has to be reconstructed, 
which is not what happens, by definition, in inclusive production studies. 
Alternatively, it is in principle possible to adopt techniques commonly applied in cases 
where only a partial reconstruction of the relevant process is practically allowed. 
This option foresees, for example, the definition of simulated template distributions 
corresponding to different natural-polarization hypotheses, 
using inputs from models detailing the mixture of $\mathrm{B} \to \jpsi \, X$ decays 
effectively contributing to the non-prompt events, 
or at least providing the distribution of the masses of the accompanying particles ($X$). 
It might be worth emphasising
that our study does not address possible measurements of the natural polarization, 
but rather focuses on the experimental outcome of a polarization measurement 
where prompt and non-prompt events, whether discriminated or not from one another, 
undergo the same ``inclusive'' analysis technique, 
independently of any model inputs.
This analysis approach is, in fact, the one followed in all \jpsi polarization measurements published so far. 
Also the existing NRQCD predictions of the non-prompt \jpsi polarization 
directly provide the ``inclusively observable'' non-prompt \jpsi polarization, 
defined as a counterpart to the prompt one, for a given collision system and energy, 
and not the natural polarization of the \jpsi, 
which would be observed using the B rest frame as ``laboratory'' 
and would have no direct dependence on the collision type and energy 
(except for a possible variation of the mixture of B meson species).

Our analysis started with a detailed description of the kinematics 
of the production of \jpsi mesons in decays of $J = 0$ particles.
In certain experimental and kinematic conditions,
such decays represent an extreme example of ``polarization smearing'', 
potentially leading to the observation of unpolarized production.
The \jpsi is, in fact, intrinsically polarized along 
the direction of its emission in the B rest frame (the cHX frame),
being, in particular, fully longitudinally polarized
when $X$ is a single $J=0$ particle, such as a kaon.
However, if only the \jpsi decay is observed, 
without reconstruction of the $\mathrm{B} \to \jpsi \, X$ step, 
the dilepton distribution is necessarily referred to the directions of the colliding beams, 
using, e.g., the HX polarization frame. 
Given that the \jpsi is emitted isotropically in the B rest frame, 
in its own rest frame the directions of the HX and cHX axes 
are distributed in a spherically uniform way with respect to one another, 
leading, in principle, to a fully smeared dilepton distribution with respect to the HX axis.

The measurement process disrupts the spherical symmetry of the smearing, 
by sculpting the distribution of the B-frame emission angles, $\Theta$ and $\Phi$. 
If the measurement is made in bins of \jpsi \pt, say,
then the $\cos \Theta$ distribution ceases to be uniform and 
assumes a shape that depends on the slope of the \pt distribution within each considered \pt interval. 
The impact of this effect increases with the mother-daughter mass difference:
$\psi$(2S) mesons from B decays should have a smaller polarization magnitude 
than the analogous \jpsi mesons, and 
an almost full smearing is expected in the (kinematically analogous) decay 
$\chi_{c0} \to \jpsi \, \gamma$.

When additional criteria are applied to select the event sample,  
stronger kinematic modulations are expected for the dilepton polarization parameters. 
For example, the minimum-\pt requirements on the decay leptons 
further sculpt the $\cos \Theta$ distributions,
which are not observed and, hence, cannot be corrected for the lepton acceptance, 
an effect that leads to an increase of the observed anisotropies. 
The selection criteria must, therefore, 
be explicitly reported together with the measurement's 
domain of all involved kinematic variables.
%
Clearly, a four-dimensional analysis of the two-step angular distribution, 
taking into account acceptance correlations between the 
$(\cos\Theta, \Phi)$ and $(\cos\vartheta, \varphi)$ variables, 
would be immune to the smearing effects and, furthermore,
to the non-physical anisotropies created by the event selections 
(which could be corrected for).
This approach would determine the full natural polarization of the \jpsi 
with respect to the B rest frame and can be adopted in exclusive decay analyses, 
where, e.g., the accompanying kaon is fully reconstructed.
Obviously, this is not a possible option when the polarization is measured
using an event sample of non-prompt \jpsi mesons.

We conclude that polarization measurements of non-prompt \jpsi mesons,
where only the dilepton angular degrees of freedom are considered,
must be reported together with a detailed and reproducible definition of the 
(single-lepton and dilepton) phase space window where the analysis is made.
Two measurements (or a measurement and a theory calculation)
can only be reliably compared if those kinematical constraints are accounted for.
The exact comparison between two results may even require 
detailed simulations of the experimental conditions, 
including any other event selections that may have an effect on the $\cos\Theta$ distribution.

As a result of this study, we found that
the non-prompt \jpsi dilepton decay distribution tends to be significantly anisotropic
if the sample is dominated by two-body decays of the kind $\mathrm{B} \to \jpsi \, \mathrm{K}$.
Multi-body processes, including, among others, 
the cascade chains with intermediate $\chi_c$ or $\psi$(2S) mesons, dilute the overall polarization.

In particular, a more pronounced longitudinal polarization, in the HX frame,
should be measured when using a non-prompt \jpsi event sample with a stronger two-body component.
More quantitatively, for \pt exceeding $\sim$\,20\,GeV,
the $\lth^{\mathrm{HX}}$ values characterizing the two-body and multi-body topologies
should differ by around 0.2, a difference large enough to be resolved by 
analyses of LHC data.
The potential discrimination between the two kinds of topologies is interesting 
because two-body decays reasonably tend to be dominated by colour-singlet processes, 
while multi-body ones should include processes producing a colour-octet \QQbar 
state~\cite{Beneke:1998ks,Beneke:1999gq,BaBar}. 
This means that measurements of non-prompt \jpsi polarization can provide valuable information 
on the charmonium production mechanism, 
complementary to the existing prompt production results.
For their correct interpretation, 
it is important that the methodological aspects discussed in this article are taken into account,
both in the data analyses and in the theoretical interpretations.

\bigskip
The authors acknowledge support from 
Funda\c{c}\~ao para a Ci\^encia e a Tecnologia, Portugal, 
under contract CERN/FIS-PAR/0010/2019

\bibliographystyle{cl_unsrt}
\bibliography{NonPrompt-Jpsi-polarization}{}

\end{document}